\documentclass[11pt,a4paper]{article}
\pdfoutput=1
\usepackage{jcappub}
\usepackage{bm}
\usepackage{bbm}
\usepackage{amsmath}

\newcommand{\llangle}{\hbox{${\big\langle}\kern-3.5pt{\big\langle}$}}
\newcommand{\rrangle}{\hbox{${\big\rangle}\kern-3.6pt{\big\rangle}$}}

\def\Xint#1{\mathchoice
   {\XXint\displaystyle\textstyle{#1}}%
   {\XXint\textstyle\scriptstyle{#1}}%
   {\XXint\scriptstyle\scriptscriptstyle{#1}}%
   {\XXint\scriptscriptstyle\scriptscriptstyle{#1}}%
   \!\int}
\def\XXint#1#2#3{{\setbox0=\hbox{$#1{#2#3}{\int}$}
     \vcenter{\hbox{$#2#3$}}\kern-.5\wd0}}

\def\dashint{\Xint-}

\renewcommand\){\right)}
\renewcommand\[{\left[}

\newcommand{\GF}{G_{\rm F}}
\newcommand{\bpartial}{{\bm\partial}}
\newcommand{\bp}{{\bf p}}
\newcommand{\bx}{{\bf x}}
\newcommand{\br}{{\bf r}}
\newcommand{\bv}{{\bf v}}

\newcommand{\bK}{{\bf K}}

\newcommand{\uc}{u_{\rm c}}
\newcommand{\wc}{w_{\rm c}}
\newcommand{\mc}{\tilde\mu_{\rm c}}
\newcommand{\sH}{{\sf H}}
\newcommand{\cH}{{\cal H}}
\newcommand{\cG}{{\cal G}}
\newcommand{\cP}{{\cal P}}

\newcommand{\half}{{\textstyle\frac{1}{2}}}

\long\def\exclude#1{}

\begin{document}
\subheader{\hfill MPP-2019-120}

\title{Fast Neutrino Flavor Conversion:
Collective Motion vs.\ Decoherence}

\author{Francesco~Capozzi,}
\author{Georg~Raffelt}
\author{and Tobias~Stirner}

\affiliation{Max-Planck-Institut f\"ur Physik (Werner-Heisenberg-Institut),\\
  F\"ohringer Ring 6, 80805 M\"unchen, Germany}

\emailAdd{capozzi@mpp.mpg.de}
\emailAdd{raffelt@mpp.mpg.de}
\emailAdd{stirner@mpp.mpg.de}

\abstract{In an interacting neutrino gas, flavor coherence becomes
  dynamical and can propagate as a collective mode.  In particular,
  tachyonic instabilities can appear, leading to ``fast flavor
  conversion'' that is independent of neutrino masses and mixing
  angles. On the other hand, without neutrino-neutrino interaction, a
  prepared wave packet of flavor coherence simply dissipates by
  kinematical decoherence of infinitely many non-collective modes. We
  reexamine the dispersion relation for fast flavor modes and show
  that for any wavenumber, there exists a continuum of non-collective
  modes besides a few discrete collective ones. So for any initial
  wave packet, both decoherence and collective motion occurs, although
  the latter typically dominates for a sufficiently dense gas.  We
  derive explicit eigenfunctions for both collective and
  non-collective modes. If the angular mode distribution of
  electron-lepton number crosses between positive and negative values,
  two non-collective modes can merge to become a tachyonic collective
  mode. We explicitly calculate the interaction strength for this
  critical point. As a corollary we find that a single crossing always
  leads to a tachyonic instability. For an even number of crossings,
  no instability needs to occur.}

\maketitle

\section{Introduction}
\label{sec:introduction}

Beginning fifty years ago with the ``solar neutrino problem'' of an
apparent $\nu_e$ flux deficit from the Sun and twenty years ago with the
up-down asymmetry of atmospheric neutrinos, the issue of flavor
conversion has developed to a main-stream field of experimental and
theoretical investigation. One intruiging ingredient is the refractive
effect on neutrino propagation by a background medium
\cite{Wolfenstein:1977ue} that can lead to large flavor conversion
even for small mixing angles (MSW effect \cite{Mikheev:1986gs}). It
remains forever fascinating that for many cases of interest, the
vacuum oscillation frequency $\Delta m^2/2E$ and the matter potential
$\sqrt{2}\GF n_e$ are similar, one depending on the small neutrino
mass differences and the other on their weak interaction strength.

One theoretical frontier that remains somewhat unsettled is neutrino
propagation in a medium consisting of other neutrinos
\cite{Pantaleone:1992eq, Samuel:1993uw, Duan:2006an, Hannestad:2006nj,
  Duan:2009cd, Duan:2010bg, Chakraborty:2016yeg}.  In particular, this
question concerns collapsing stars, often leading to core-collapse
supernovae, and neutron-star mergers. Neutrinos play a dominant
dynamical role in such events and their flavor composition likely
shapes nucleosynthesis in the concomitant neutrino-driven matter
outflows. Moreover, interpreting the flavor composition of the
neutrino signal from the next nearby supernova will depend on a better
theoretical understanding of neutrino-neutrino refraction in the
source region.

In the mean-field approximation, neutrino radiation is represented by
density matrices $\varrho_{\bp}(t,\bx)$ in flavor space for each
momentum mode $\bp$, where the diagonal entries are the usual
occupation numbers, while the off-diagonal elements encode flavor
coherence. In the ultrarelativistic limit, the space-time evolution
follows from the kinetic equation~\cite{Dolgov:1980cq,
  Rudzsky:1990, Sigl:1992fn, Sirera:1998ia, Yamada:2000za,
  Cardall:2007zw, Vlasenko:2013fja,Hansen:2016klk,
  Stirner:2018ojk,Richers:2019grc}
\begin{equation}\label{eq:hamiltonian-1}
 (\partial_t+\bv\cdot\bpartial_\br)\,\varrho_{\bp}=
  -i[\sH_\bp,\varrho_\bp]+{\cal C}(\varrho_\bp)\,,
\end{equation}
where the neutrino velocity vector $\bv=\bp/|\bp|$ is taken to be a
unit vector describing the direction of motion. The left-hand side
(lhs) is the advection term provided by the Liouville operator and
describes free streaming, assuming we ignore momentum changes by
coherent forces and notably gravitational bending of trajectories.
The collision term on the right-hand side (rhs) accounts for
scattering and emission or absorption.  Coherent flavor evolution is
governed the Hamiltonian matrix $\sH_\bp$ that depends on neutrino
masses and the flux densities of background particles including other
neutrinos.

Ignoring coherent flavor conversion means replacing $\sH_\bp$ by the
unit matrix, in which case flavor evolution can arise only by
collisions such as pair annihilation and charged-current processes
that are encoded in the collision term.  In numerical supernova (SN)
simulations, this assumption is always used and justified by the large
matter effect which effectively demixes neutrinos, i.e., propagation
eigenstates are very nearly identical to flavor eigenstates. Coherent
flavor evolution, e.g.\ by MSW conversion, has usually been studied by
means of post-processing. It remains to be understood if it is
necessary to incorporate coherent flavor evolution in numerical SN
simulations to obtain reliable results regarding the
neutrino-driven explosion
mechanism and nucleosynthesis.

One recent focus of attention concerns ``fast flavor conversion,''
i.e., nontrivial dynamical solutions of
equation~\eqref{eq:hamiltonian-1} that do not depend on the vacuum
oscillation frequency and would exist even for vanishing neutrino
masses. Some initial off-diagonal seed in the $\varrho_{\bp}(t,\bx)$
distribution could lead to self-induced flavor conversion
\cite{Sawyer:2005jk, Sawyer:2008zs, Sawyer:2015dsa,
  Chakraborty:2016lct, Dasgupta:2016dbv, Dasgupta:2017oko,
  Abbar:2017pkh}.  The relevant length scale would be governed by a
typical neutrino-neutrino refractive energy $\mu=\sqrt2\GF
n_{\nu\bar\nu}$ that far exceeds the vacuum oscillation frequency. So
these effects could be important on short length scales and modify
neutrino flavor evolution near the neutrino decoupling region in a
SN core or in neutron-star binary mergers.

Several questions need answering. What are the required conditions for
the neutrino gas to support fast collective modes? If these conditions
are satisfied, what exactly triggers these modes? And if they are
started, will unstable modes indeed reach the nonlinear phase and
cause tangible effects relevant for the physics of compact objects?

The first of these questions can be nicely addressed in terms of a
normal-mode analysis of the linearised version of the kinetic equation
\cite{Banerjee:2011fj, Abbar:2015fwa, Izaguirre:2016gsx,
  Capozzi:2017gqd, Capozzi:2018clo, Airen:2018nvp, Morinaga:2018aug,
  Azari:2019jvr, Yi:2019hrp}. The crucial ingredient for supporting
possible fast tachyonic instabilities appears to be the angle distribution
of the electron lepton number (ELN) carried by neutrinos that seems to
require a ``crossing'' between positive and negative values, although
the concept of the ELN distribution would need to be formulated more
generally if the other flavors, notably the muon neutrinos, also
carry a flavor lepton number flux.  Several groups have studied the
ELN angle distribution in various types of SN and binary merger
simulations to look for astrophysical environments that would show such
crossings and thus would support
fast-flavor conversion \cite{Tamborra:2017ubu, Wu:2017qpc,
  Wu:2017drk, Abbar:2018shq, Capozzi:2018rzl, Shalgar:2019kzy}.

We here return to more basic questions of the
normal-mode analysis for the linearised version of
equation~\eqref{eq:hamiltonian-1} to clarify the emergence
of collective modes from the multitude of non-collective ones that
exist in the absence of neutrino refraction and otherwise co-exist with
collective modes. In the ``fast
flavor limit'' of vanishing neutrino masses, and ignoring neutrino-neutrino
refraction, any prepared initial condition of flavor coherence would be
dissipated by the multi-directional motion of the various neutrino modes
given by the advection term. In other words, the dispersion relation
corresponding to the kinetic equation should always include all modes,
collective and non-collective ones, even if it is only the former that
ultimately are of interest for fast flavor conversion.

To this end we begin in section~\ref{sec:EOM} with the linearised equation of motion
and reduce its dimensionality to an axially symmetric system that offers
the simplest sandbox for our demonstration.
In section~\ref{sec:non-coll-modes} we explicitly consider the non-collective
modes and derive the explicit form of their singular eigenfunctions, notably
in the presence of nonvanishing neutrino-neutrino refraction and we show how
collective modes emerge from the continuum of non-collective ones.
As a somewhat unexpected bonus, the explicit identification
of the non-collective modes allows us in section~\ref{sec:crossings}
to identify the critical points where a crossed ELN angle spectrum
produces a branch point for a tachyonic solution to appear from the
coalescence of two non-collective modes. In section~\ref{sec:decoherence}
we return to our original goal and show, in a simple example, how
collective motion takes over from the non-collective modes with
increasing interaction strength. Section~\ref{sec:conclusions} is finally given
over to conclusions.

\section{Linearised equation of motion}

\label{sec:EOM}

\subsection{Kinetic equation for fast flavor modes}

The starting point for our study is the kinetic equation~\ref{eq:hamiltonian-1}
where we neglect
the collision term and work in the ``fast flavor limit'' where
neutrino masses and mixing are ignored. Moreover, the background
medium is taken to be homogeneous, isotropic, and stationary, in which
case the matter effect can be ``rotated away.'' Therefore, the
Hamiltonian matrix on the right-hand side (rhs) includes only
neutrino-neutrino interactions and thus has the form
\begin{equation}
  \sH_\bp=\sqrt{2}\GF \int\frac{d^3\bp'}{(2\pi)^3}\,(1-\bv\cdot\bv')
  (\varrho_{\bp'}-\bar\varrho_{\bp'})\,,
\end{equation}
where $\bar\varrho_{\bp}$ is the density matrix for antineutrinos.

We linearise the kinetic equation, so we can limit our
discussion to a two-flavor system consisting of $\nu_e$ and some
other flavor $\nu_x$. We write the density matrices as
\begin{equation}\label{eq:s-define}
  \varrho_\bp=\frac{f_{\nu_e,\bp}+f_{\nu_x,\bp}}{2}\,\mathbbm{1}
  +\frac{f_{\nu_e,\bp}-f_{\nu_x,\bp}}{2}
  \begin{pmatrix}s_\bp&S_\bp\\S_\bp^*&-s_\bp\end{pmatrix}\,,
\end{equation}
where $s_\bp$ is real, $S_\bp$ is complex, and $s_\bp^2+|S_\bp|^2=1$.
To linear order $s_\bp=1$, so we ask only for the space-time
evolution of $S_\bp$ which holds the information about flavor
coherence. Rotating away also the diagonal neutrino-neutrino matter
effect, the linearised EOM is
\begin{equation}\label{eq:EOM5}
  i(\partial_t+\bv\cdot\bpartial_\br)\,S_{\bp}=
  -\sqrt{2}\GF\int\frac{d^3\bp}{(2\pi)^3}\,(1-\bv\cdot\bv')
  \(S_{\bp'}f_{\bp'}-\bar S_{\bp'}\bar f_{\bp'}\).
\end{equation}
An analogous equation applies to the antineutrino flavor coherence
$\bar S_\bp$. Here we use the difference spectra
$f_\bp=(f_{\nu_e,\bp}-f_{\nu_\mu,\bp})$ and $\bar
f_\bp=(f_{\bar\nu_e,\bp}-f_{\bar\nu_\mu,\bp})$.

The EOM does not depend on neutrino energy and is the same for neutrinos
and antineutrinos. Therefore, assuming identical initial conditions, we may
integrate over energies and sum over neutrinos and antineutrinos to obtain an
EOM that depends only on the direction $\bv$ of a given mode,
\begin{equation}\label{eq:EOM6}
  i(\partial_t+\bv\cdot\bpartial_\br)\,S_{\bv}=
  -\sqrt{2}\GF\int\frac{d\bv'}{4\pi}\,(1-\bv\cdot\bv')\,
  g_{\bv'}\,S_{\bv'}\,.
\end{equation}
The integration is over all directions $\bv$, i.e., over the unit
sphere in the space of velocities.
The angle distribution of the effective density of electron-lepton number (ELN) is
\begin{equation}\label{eq:gv-define}
  g_\bv=\int_{0}^{\infty}\frac{E^2 dE}{2\pi^2}\,
  \bigl(f_{\nu_e,\bp}-f_{\bar\nu_e,\bp}-f_{\nu_x,\bp}+f_{\bar\nu_x,\bp}\bigr)
\end{equation}
with $\bp=E\,\bv$.

\subsection{Axial symmetry}

We study a restricted class of backgrounds and solutions, where the neutrino angle
distribution is axially symmetric relative to some direction $\br$ that could be
the radial direction in the supernova context. Moreover, we consider only those
flavor modes that have wave vectors along that same direction, and we assume that
the solution itself is axially symmetric, ignoring those modes that spontaneously
break axial symmetry. So finally we consider the 1+1 dimensional problem
\begin{equation}\label{eq:EOM7}
  i(\partial_t+u\partial_r)\,S_{u}=
  -\mu\int_{-1}^{+1}du'\,(1-uu')\,G_{u'}\,S_{u'}\,,
\end{equation}
where $u=\cos\theta$ is the velocity component along the symmetry direction $\br$,
and the ELN angular distribution and effective interaction strength are
\begin{subequations}
\begin{eqnarray}\label{eq:Gv-define}
  G_u&=&\frac{1}{n_{\nu_e}+n_{\bar\nu_e}}\int_{-\pi}^{+\pi}
  \frac{d\varphi}{4\pi} \int_{0}^{\infty}\frac{E^2 dE}{2\pi^2}\,
  \bigl(f_{\nu_e,\bp}-f_{\bar\nu_e,\bp}-f_{\nu_x,\bp}+f_{\bar\nu_x,\bp}\bigr)\,,
\\[2ex]
  \mu&=&\sqrt{2}\,\GF(n_{\nu_e}+n_{\bar\nu_e})\,.
\end{eqnarray}
\end{subequations}
In this way, $G_u$ is a dimensionless function with values of the order of unity,
whereas $\mu$ (units of energy) is an effective interaction strength
between neutrinos. We have arbitrarily normalised
these quantities to the sum of the $\nu_e$ and $\bar\nu_e$ densities, but of course other
definitions are possible. There is no entirely natural way to
normalise the dimensionless angle distribution.

\subsection{Dispersion relation}
\label{sec:dispersion}

As equation~\eqref{eq:EOM7} is linear, we can solve it in Fourier space and consider
solutions of the form
\begin{equation}\label{eq:eigen-1}
  S_u(t,r)=Q_u(\Omega,K)\,e^{-i(\Omega t-Kr)}\,,
\end{equation}
where the eigenfunction obeys the equation
\begin{equation}\label{eq:EOM8}
  (\Omega-u K)\,Q_u(\Omega,K)
  =-\mu\int_{-1}^{+1}du'\,(1-uu')\,G_{u'}\,Q_{u'}(\Omega,K)\,.
\end{equation}
For a given angle distribution $G_u$ and interaction strength $\mu$ we can
solve this equation and find, for given $K$, the corresponding $\Omega$
as well as the eigenfunction $Q_u$ for $(\Omega,K)$.

The solution is trivial in the absence of interactions ($\mu=0$). For given
$K$, any $\Omega=w K$ with $-1\leq w\leq +1$ is a solution with the eigenfunctions
$Q_u(\Omega,K)=\delta(u-w)$. Here $w=\Omega/K$ has the interpretation of the phase
velocity for the given wave, which here is less than the speed of light, commensurate
with the picture that these modes describe the flavor coherence of a single neutrino
mode (or rather a cone of modes)
$u=\cos\theta$ relative to the radial direction in a supernova and thus
travels with $-1\leq u\leq+1$ along that direction. If we were to set up
a wave, or a wave packet, over many modes, it would quickly dissipate
by decoherence because of the continuum of frequencies with which it
would oscillate. Notice that these modes are not dynamical---the oscillation
with frequency $\Omega$ seen by an observer in the laboratory frame arises
because a wave with wave number $K$ simply drifts by. This is a purely
kinematical effect. One focus of our paper is to understand what happens
to these modes once the interaction is turned on.

First, however, we recapitulate the dispersion relation for collective modes that
appear in the presence of a nonvanishing $\mu$ and that involve all
directions (or cones) $u$ simultaneously, in
contrast to the kinematical modes just described.
To this end we observe that the rhs of equation~\eqref{eq:EOM8} is of the
form $a+b u$ with unknown coefficients $a$ and $b$. Therefore, the eigenfunctions
must have the form
\begin{equation}\label{eq:eigen-2}
  Q_u=\frac{a+bu}{w-u}
\end{equation}
so long as $w=\Omega/K$ is outside of the interval $[-1,+1]$
and thus the denominator does not become singular for
$u$ within this interval. In other words, for given $K$, this ansatz applies
to real $\Omega$ ``outside of the light cone'' with phase velocity $|w|>1$
or to unstable modes where $\Omega$ has an imaginary part.
Inserting this ansatz on both sides in equation~\eqref{eq:EOM8} yields
\begin{equation}\label{eq:EOM9}
  a+bu=-\frac{\mu}{K}\int_{-1}^{+1}du'\,\frac{(1-uu')(a+bu')}{w-u'}\,G_{u'}\,.
\end{equation}
This equality must apply for any $u$ and thus represents two equations,
one consisting of the terms that are independent of $u$ and the other
linear in $u$. Therefore, after dropping the prime in the integration
variable $u'$, one finds
\begin{equation}\label{eq:EOM10}
  \begin{pmatrix}
    +K & \\
    & -K
  \end{pmatrix}\begin{pmatrix}
    a \\
    b
  \end{pmatrix}=-\mu\left[
  \int_{-1}^{+1}du\,\frac{G_{u}}{w-u}
  \begin{pmatrix}
    1 & u \\
    u&u^2
  \end{pmatrix}\right]
  \begin{pmatrix}
    a \\
    b
  \end{pmatrix}.
\end{equation}
This equation has nontrivial solutions for $a$ and $b$ if the determinant vanishes,
\begin{equation}\label{eq:determinant-1}
  \left\|\,\begin{pmatrix}
             K &     \\
                &  -K
           \end{pmatrix}+\mu
\begin{pmatrix}
             \langle 1\rangle_w & \langle u\rangle_w   \\
             \langle u\rangle_w  & \langle u^2\rangle_w
           \end{pmatrix}
\right\|=0\,,
\end{equation}
where we use the notation
\begin{equation}\label{eq:G-integrals}
  \langle u^n\rangle_w=\int_{-1}^{+1}du\,G_u\,\frac{u^n}{w-u}\,.
\end{equation}
This is a quadratic equation with two solutions
\begin{equation}\label{eq:K-solutions}
K_w=-\mu\frac{\kappa_w\pm\sqrt{\Delta_w}}{2}\,,
\end{equation}
where $\kappa_w = \bigl\langle 1- u^2\bigr\rangle_w$ and
$\Delta_w = \bigl\langle (1-u)^2\bigr\rangle_w\,\bigl\langle (1+u)^2\bigr\rangle_w$.
For those values of $w$ where $\Delta_w\geq0$, we find the parametric solutions
$(\Omega,K)=(w K_w,K_w)$ of propagating waves.

In the limit $w\to\pm\infty$, where we can neglect $u$ in the denominator
of equation~\eqref{eq:G-integrals}, we find
$\langle u^n\rangle_w\to \cG_n/w$, where we use the moments of the $G$-distribution
\begin{equation}\label{eq:G-moments}
  \cG_n=\int_{-1}^{+1}du\,G_u u^n\,.
\end{equation}
In the $w\to\infty$ limit, $\kappa_w$ and $\Delta_w$ scale with $1/w$ so that $K_w\to 0$,
i.e., this is the case of vanishing wave number. The corresponding $\Omega_w$ involve
an extra power of $w$ and become
\begin{equation}
\Omega_{w\to\pm\infty}=
-\mu\,\frac{\cG_0-\cG_2\pm\sqrt{\left(\,\cG_0+\cG_2\right)^2-4\cG_1^2}}{2}\,.
\end{equation}
We find the same result if we consider the homogeneous
case $K=0$ from the start and solve the eigenvalue equation for $\Omega$, as it was done 
previously in reference~\cite{Dasgupta:2018ulw}.

\begin{figure}[ht]
  \hbox to\textwidth{\hfil
  \includegraphics[scale=0.6]{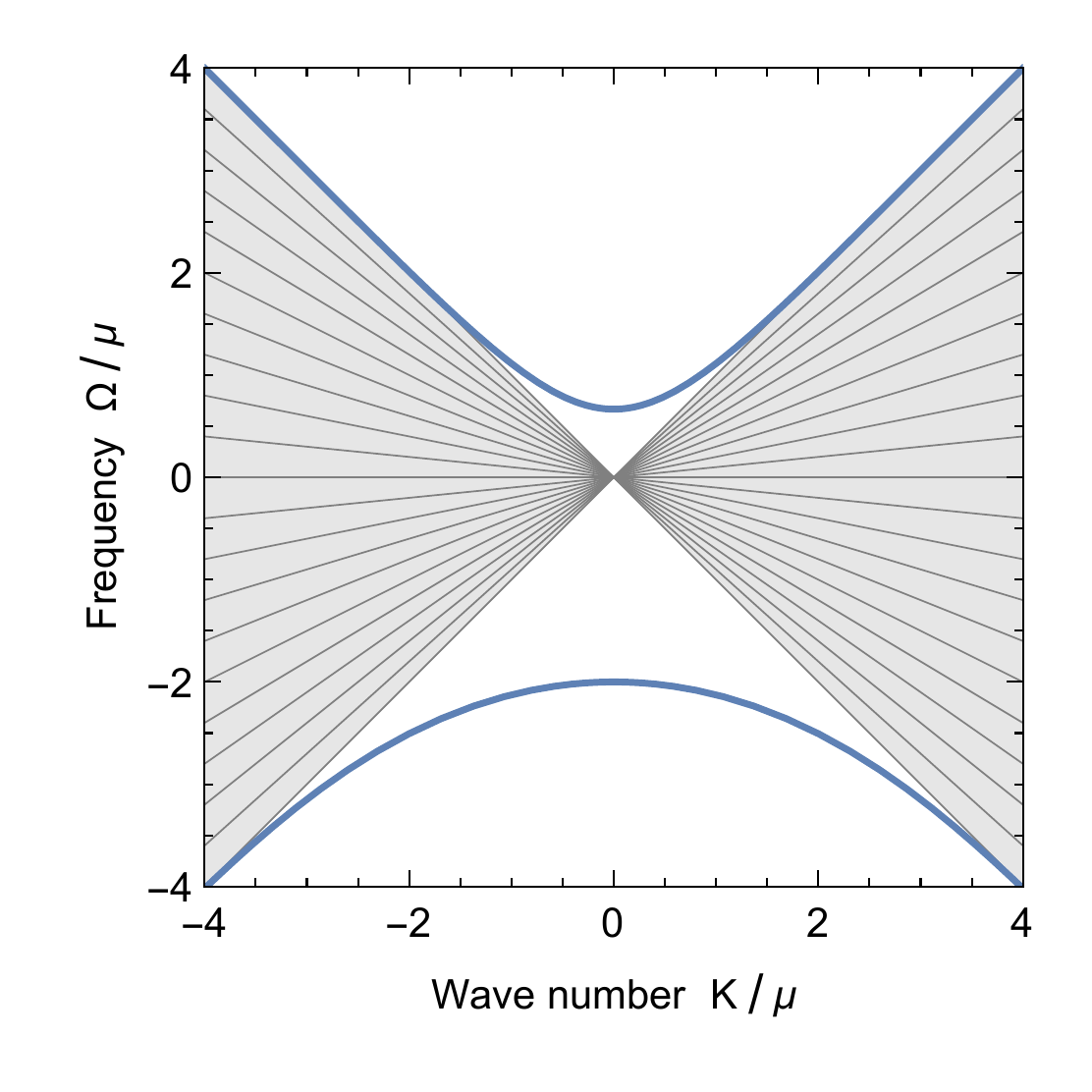}\hskip20pt
  \includegraphics[scale=0.6]{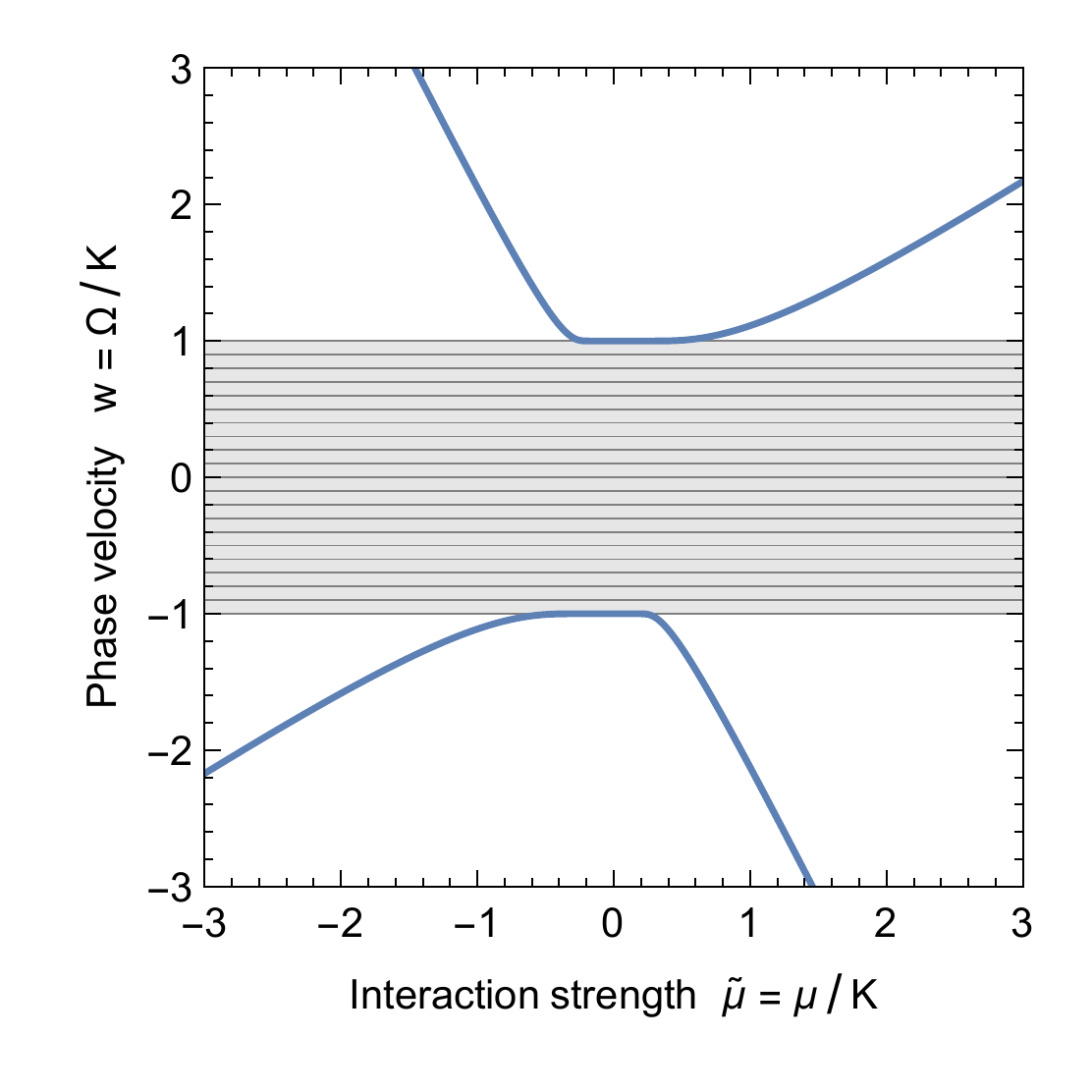}\hfil}
  \caption{Dispersion relation $\Omega(K)$ for flavor modes in an isotropic neutrino gas.
  {\em Thick blue lines:} Collective modes.  {\em Thin gray lines:}
  Examples for non-collective modes that provide dense coverage under
  the light cone.
  {\rm Right panel:} Phase velocity $w=\Omega/K$ as a function of
  the effective interaction strength $\tilde\mu=\mu/K$. Notice that
  the upper left quadrant of the right panel corresponds to the lower left one
  of the left panel because $w=\Omega/K$ is positive when $\Omega$ and $K$
  are both negative.}
  \label{fig:isotropic-dispersion}
\end{figure}

If  $G_u$ is either positive or negative on the interval $-1\leq u\leq+1$
(no crossing between positive and negative values)
and for $|w|>1$, the averages in the definition of $\Delta_w$ have equal signs so that
$\Delta_w\geq 0$. Therefore, in the absence of $G_u$-crossings we find two real branches of the
dispersion relation and no tachyonic instabilities.
For the simplest case of an isotropic medium with $G_u=1$ we show the two branches
of the collective dispersion relation in the left panel of figure~\ref{fig:isotropic-dispersion}
as thick blue lines. Under the light cone (gray shaded region) we show in addition a grid of
non-collective modes, assuming they are not modified by the interaction. They provide dense
coverage under the light cone ($-1<w<+1$).

We will find it useful to display the same information in the form of the right panel of
figure~\ref{fig:isotropic-dispersion}, where on the vertical axis we show
the phase velocity $w=\Omega/K$ for a given mode as a function of $\tilde\mu=\mu/K$ on the
horizontal axis. The light cone is now simply the shown gray horizontal band and the
non-collective modes are horizontal lines, assuming they are not modified
by the interaction. The case of vanishing interaction $\tilde\mu=0$ is at the
center of the plot. It corresponds to the asymptotic
solution $K\to\pm\infty$ in the left panel. While $\mu$ is defined as a positive
parameter, $\tilde{\mu}$ can be both positive and negative.

\subsection{Schr\"odinger-like equation}

The representation of the right panel of figure~\ref{fig:isotropic-dispersion}
corresponds to a somewhat different perspective on the dispersion relation. The only
dimensionful parameter of our problem is the interaction strength $\mu$, so all frequencies and
wave numbers can be expressed in units of $\mu$. However, we can also imagine that we pick
one fixed wave number $K$ and ask for the spectrum of eigenfrequencies $\Omega$, depending
on the chosen $\mu$, and express the eigenfrequencies in units of $K$. In this way,
what was the phase velocity $w=\Omega/K$ can be seen as the eigenfrequency in units
of $K$ and we ask how it behaves as a function of the interaction strength $\tilde{\mu}$.
In this picture, equation~\eqref{eq:EOM8} is
\begin{equation}\label{eq:EOM8a}
  (w-u)\,Q_{w,u}
  =-\tilde\mu\int_{-1}^{+1}du'\,(1-uu')\,G_{u'}\,Q_{w,u'}\,,
\end{equation}
where the chosen wave number is hidden in $\tilde{\mu}$. Returning to an equation
for the time evolution, this corresponds to
\begin{equation}\label{eq:EOM8b}
  i\partial_t S=\cH(S)\,,
\end{equation}
where dimensionless time is measured in units of $1/|K|$, but otherwise $K$ does not appear.
Here $S$ is a time-dependent complex function of $u$ on the interval $-1\leq u\leq+1$ and
the linear operator on the rhs is
\begin{equation}\label{eq:Hamiltonian}
  \cH(S)\big|_u=u\,S_u(t)-\tilde\mu\int_{-1}^{+1}du'\,(1-uu')\,G_{u'}\,S_{u'}(t)\,.
\end{equation}
So the right panel of figure~\ref{fig:isotropic-dispersion} shows the
eigenvalues of the operator ${\cal H}$ as function of~$\tilde{\mu}$.
For vanishing interaction, the spectrum of eigenvalues is the continuum
$-1\leq w\leq +1$, whereas for nonvanishing $\tilde{\mu}$, in addition two
discrete collective modes emerge.

We can think of equation~\eqref{eq:EOM8b} as a Schr\"odinger equation with Hamiltonian
$\cH$ and eigenvalues $w$. However, unless $G_u={\rm const.}$,
the real operator $\cH$ is not symmetric and thus not Hermitean
and indeed can have complex eigenvalues. The latter are, of course, the main interest
concerning collective modes.
Our linearised EOMs are not complete---we have ignored the evolution of $s_{\bp}$, so the evolution
of our sub-system need not be unitary.

We should solve the eigenvalue equation and find the energies $w$ as a function of the
interaction strength $\tilde\mu$. For real $w$ we can turn this question around and find
\begin{equation}\label{eq:mu-value}
  \tilde{\mu}=-\frac{2}{\kappa_w\pm\sqrt{\Delta_w}}\,,
\end{equation}
where we have used the results of section~\ref{sec:dispersion}.
It is then straightforward to find the coefficients $a$ and $b$ for the
normalised eigenfunctions of the form~\eqref{eq:eigen-2}, although the
expressions get cumbersome.

\section{Non-collective modes}

\label{sec:non-coll-modes}

\subsection{Eigenfunctions}

To determine the non-collective modes (the propagating ones within the light cone),
we expect that for any $\tilde{\mu}$ the spectrum of eigenfrequencies densely fills
the interval $-1<w<+1$. So for a given $\tilde{\mu}$ we know the spectrum of
non-collective eigenvalues and all that is missing is the form of the eigenfunctions.
In the noninteracting case they are proportional to $\delta(w-u)$, so switching on
a nonvanishing $\tilde{\mu}$ should also involve a $\delta$ function. On the other hand,
it should also have the form of equation~\eqref{eq:eigen-2} by the same logic that the
rhs of equation~\eqref{eq:EOM8a} is of the form $a+b u$. If we add $-bw+bw$ in the numerator
of equation~\eqref{eq:eigen-2} we see that it is $(a+bw)/(w-u)+b$ or $A/(w-u)+B$ in terms
of other parameters $A$ and $B$. So overall we propose that the eigenfunctions should
be of the form
\begin{equation}\label{eq:eigenfunctions-non-coll}
  Q_{w,u}=A_w\left[\frac{1}{w-u}+\alpha_w\delta(w-u)\right]+B_w
\end{equation}
with the stipulation that in equation~\eqref{eq:EOM8a}
the principal value of the integral should be taken. Or turning this
point around, the integral by itself is not defined and
can be made meaningful by the principal-part prescription
up to an unknown constant that must be fixed by the unknown coefficient
$\alpha_w$.

Inserting this ansatz on both sides of equation~\eqref{eq:EOM8a} and using
$(w-u)\delta(w-u)=0$, analogous to the noninteracting case,
we find the condition
\begin{equation}\label{eq:newcondition}
 A_w+B_w(w-u)=-\tilde\mu A_w\left[\,\dashint_{-1}^{+1}\!du'\frac{1-uu'}{w-u'} G_{u'}
 + \alpha_wG_w(1-uw)\right]-\tilde\mu B_w\int_{-1}^{+1}\!du'(1-uu')\,G_{u'}.
\end{equation}
We use $(1-uu')/(w-u')=(1-uw)/(w-u')+u$ under the integral to write
\begin{equation}
  \dashint_{-1}^{+1}\!du'\,\frac{1-uu'}{w-u'}\,G_{u'}=
 (1-uw)\,\underbrace{\dashint_{-1}^{+1}\!du'\,\frac{G_{u'}}{w-u'}}_{\textstyle{F_w}}{}
 +u\underbrace{\int_{-1}^{+1}\!du'\,G_{u'}}_{\textstyle{\cG_0}}\,.
\end{equation}
The first integral on the rhs is a certain transform of $G(u)$ on the interval
$-1\leq u\leq +1$ which we call $F(u)$ and is here to be taken at the location $w$,
whereas the second integral is the 0th moment of $G_u$ as defined in
equation~\eqref{eq:G-moments}.
So equation~\eqref{eq:newcondition} is
\begin{equation}\label{eq:newcondition-2}
 A_w+B_w(w-u)+\tilde\mu A_w\left[(1-uw)(F_w+\alpha_wG_w)+u\cG_0\right]
 +\tilde\mu B_w\left(\cG_0-u\cG_1\right)=0.
\end{equation}
By the same argument that was used earlier, this equation represents
two independent equations for the coefficients $A_w$ and $B_w$,
one from the terms that do not depend on $u$ and one from those that are linear in $u$,
\begin{equation}
  \begin{pmatrix}
    \tilde\mu(F_w+\alpha_wG_w)+1~& \tilde{\mu}\,\cG_0+w\\[1ex]
    \tilde\mu(F_w+\alpha_wG_w)w-\tilde{\mu}\,\cG_0~&\tilde{\mu}\,\cG_1+1
  \end{pmatrix}
  \begin{pmatrix}
    A_w \\[1ex]
    B_w
  \end{pmatrix}=0\,.
\end{equation}
For any nontrivial solution, the determinant of the matrix must vanish, implying
\begin{equation}\label{eq-alpha}
  \alpha_w G_w=-F_w
  -\frac{1+\tilde\mu\cG_0(\tilde\mu\cG_0+w)+\tilde\mu\cG_1}
  {\tilde{\mu}\,(1-w^2-\tilde\mu w\cG_0+\tilde\mu\cG_1)}\,.
\end{equation}
Inserting this solution in the matrix, one easily finds
\begin{equation}\label{eq:BA-ratio}
  \beta_w\equiv\frac{B_w}{A_w}=\frac{w+\tilde\mu\cG_0}{1-w^2-\tilde\mu(w\cG_0-\cG_1)}\,,
\end{equation}
leaving open the overall normalisation.

The simplest example is an isotropic system with $G_u=1$, implying
$\cG_0=2$, $\cG_1=0$, and $F_u=\log[(1+u)/(1-u)]$. In this case
the explicit coefficients are
\begin{subequations}\label{eq:alpha-isotropic}
\begin{eqnarray}
  \alpha_w&=&\log\left(\frac{1-w}{1+w}\right)-
  \frac{1+2\tilde\mu(2\tilde\mu+w)}{\tilde\mu\,(1-w^2-2\tilde\mu w)}\,,
  \\[1ex]
  \beta_w&=&\frac{w+2\tilde\mu}{1-w^2-2\tilde\mu w}\,.
\end{eqnarray}
\end{subequations}
In general, the coefficient $\alpha_w$ is fixed by
equation~\eqref{eq-alpha}, except for the special case when $G_w=0$,
i.e., at an angular crossing. In this case, $\delta(w-u)$ is the
eigenfunction, so our ansatz would not apply.  We will see later that
in this case, for certain values of $\tilde{\mu}$ a tachyonic solution
branches off.

\subsection{Normalisation}

\label{sec:normalisation}

These eigenfunctions cannot be normalised, whereas the collective
modes with $|w|>1$ have eigenfunctions that can be normalised. This is
analogous to free vs.\ atomic bound electrons. One solution is to
confine the electron to some large box instead of infinite space. In
our case, we may use discrete angles $u_i=\cos\theta_i$ with equal
spacing $\Delta u$ for the neutrino directions, which is equivalent to
using some large box for the neutrinos in the direction of our
symmetry axis, leading to quantized neutrino momenta in that
direction.  For arbitrarily small $\Delta u$, this means
$\int_{-1}^{+1} du\,f(u)=\sum_{i=1}^{N} f_i\,\Delta u$ for some
function $f(u)$ with $f_i=f(u_i)$. For the function $\delta(u)$ with
$\int du\, \delta(u)=1$, this implies $\int du\,|\delta(u)|^2=1/\Delta
u$ because the $\delta$ function becomes a Kronecker $\delta$ at one
grid point where it must be represented by the function value
$1/\Delta u$ to provide unit sum.

Next we consider the function $1/(w-u)$. Its principal part integrates
to a finite value, whereas its normalisation
$\int_{-1}^{+1}du\,(w-u)^{-2}$ diverges at $u=w$. To realise the
principal-part prescription in the discrete case, we need to fix the
grid points $u_i$ symmetrically around $w$. For illustration we use
$w=0$ and $u_i=(\half\pm i) \Delta u$ with $i$ the integer grid-point
indices.  Noting that $\sum_{i=-\infty}^{+\infty}(\half-i)^{-2}=\pi^2$
we find that $\int_{-1}^{+1}du\,(w-u)^{-2}=\pi^2/\Delta u$.

Finally we consider the normalisation of $Q_{w,u}$ given in
equation~\eqref{eq:eigenfunctions-non-coll}. We notice that in $\int
du\,|Q_{w,u}|^2$ the mixed terms either vanish or are finite, leaving
us with
\begin{equation}\label{eq:Q-normalisation}
  \int_{-1}^{+1}du\,|Q_{w,u}|^2=A_w^2\int_{-1}^{+1}du\,\left[\frac{1}{(w-u)^2}+\alpha_w^2|\delta(w-u)|^2\right]
  =\frac{A_w^2}{\Delta u}\,(\pi^2+\alpha_w^2)\,.
\end{equation}
So $B_\omega$ does not contribute to the normalisation and if all
eigenfunctions are to be normalised in the same way, we may fix
$A_w=a_w\,\sqrt{\Delta u}$ with
\begin{equation}\label{eq:A-normalisation}
  a_w=\frac{s_w}{\sqrt{\pi^2+\alpha_w^2}}
\qquad\hbox{where}\qquad
  s_w= {\rm sign}\!\left[-\tilde{\mu}\,(1-w^2-\tilde\mu w\cG_0+\tilde\mu\cG_1)\right]\,.
\end{equation}
Of course, the overall sign (or phase) of the eigenfunctions is not
fixed and we introduced $s_w$ for later convenience.
Notice that $B_w$ does not vanish
even though it does not
contribute to the overall normalisation of the singular eigenfunctions. Rather
$B_w$ is derived from $A_w$ through the ratio
$\beta_w$ given in
equation~\eqref{eq:BA-ratio}. We may write it in the form
$B_w=b_w\sqrt{\Delta u}$ and with equation~\eqref{eq:A-normalisation} one finds
\begin{equation}\label{eq:B-normalisation}
  b_w=-\beta_w\,\frac{s_w}{\sqrt{\pi^2+\alpha_w^2}}
\end{equation}
to achieve a common normalisation for all eigenfunctions.

We finally notice that $\alpha_w$ measures the relative contributions
of the functions $\delta(w-u)$ and $1/(w-u)$ to the eigenfunction, which thus may be written
in the form
\begin{eqnarray}\label{eq:phaseshift}
  Q_{w,u}&=&A\,\frac{s_w}{\sqrt{\pi^2+\alpha_w^2}}
  \left[\frac{1}{w-u}+\alpha_w\,\delta(w-u)+\beta_w\right]
\nonumber\\[2ex]
 &=&A\left[\frac{\sin\varphi_w}{\pi(w-u)}+\cos\varphi_w\,\delta(w-u)-b_w\right]\,,
\end{eqnarray}
where $A$ is now a chosen global normalisation factor, for example $A=\sqrt{\Delta u}$,
that does not depend on the eigenvalue $w$. The angle $\varphi_w$ is given by the relations
\begin{equation}\label{eq:phaseshift-2}
  \sin\varphi_w=\frac{\pi}{\sqrt{\pi^2+\alpha_w^2}}\,s_w
  \qquad\hbox{and}\qquad
  \cos\varphi_w=\frac{\alpha_w}{\sqrt{\pi^2+\alpha_w^2}}\,s_w\,.
\end{equation}
For vanishing interaction strength, one finds $\alpha_w\to-\infty$, $\varphi_w=0$,
and $B_w=0$ so that we have only the $\delta$ function.

\begin{figure}[ht]
  \hbox to\textwidth{\hfil
  \includegraphics[scale=0.52]{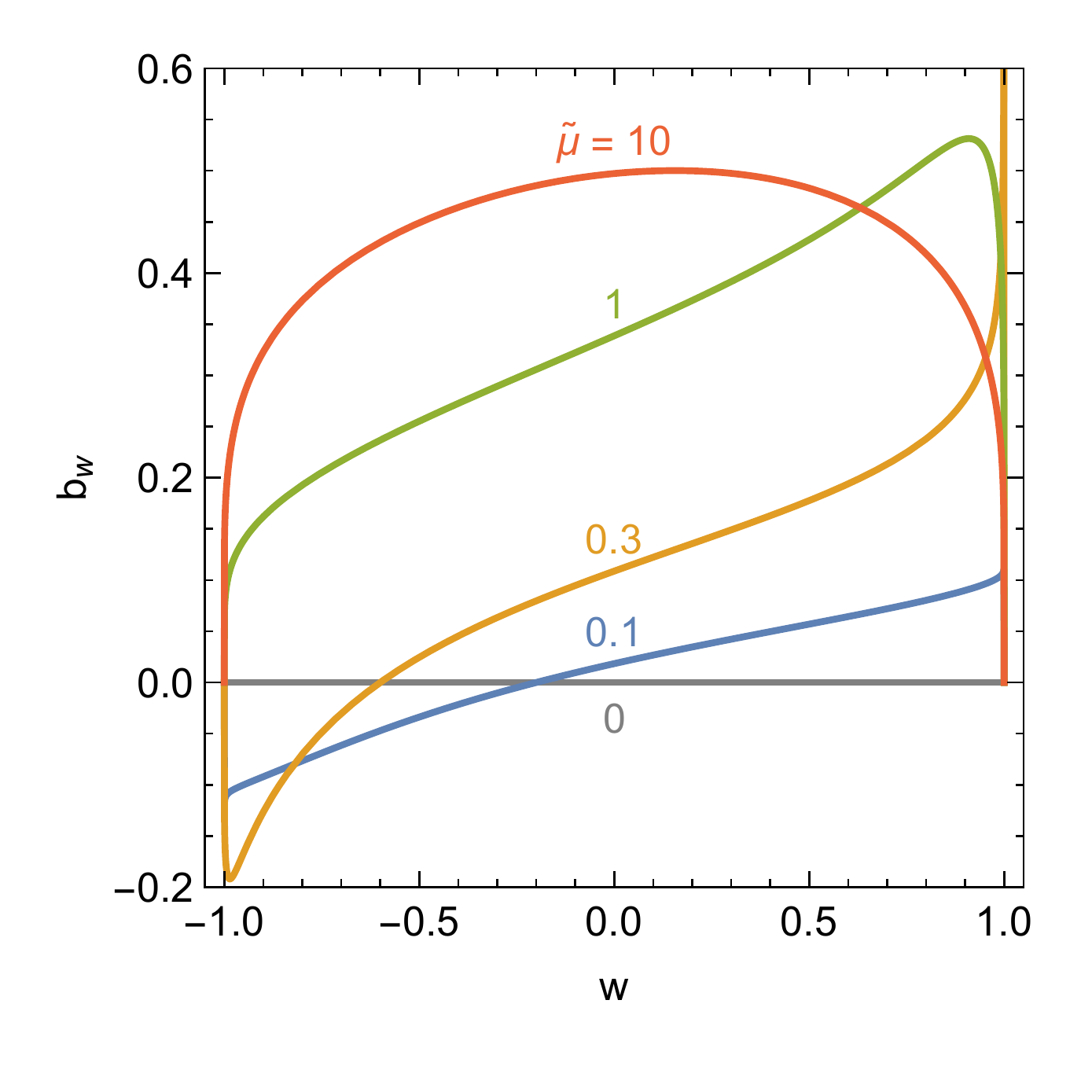}\hskip20pt
  \includegraphics[scale=0.52]{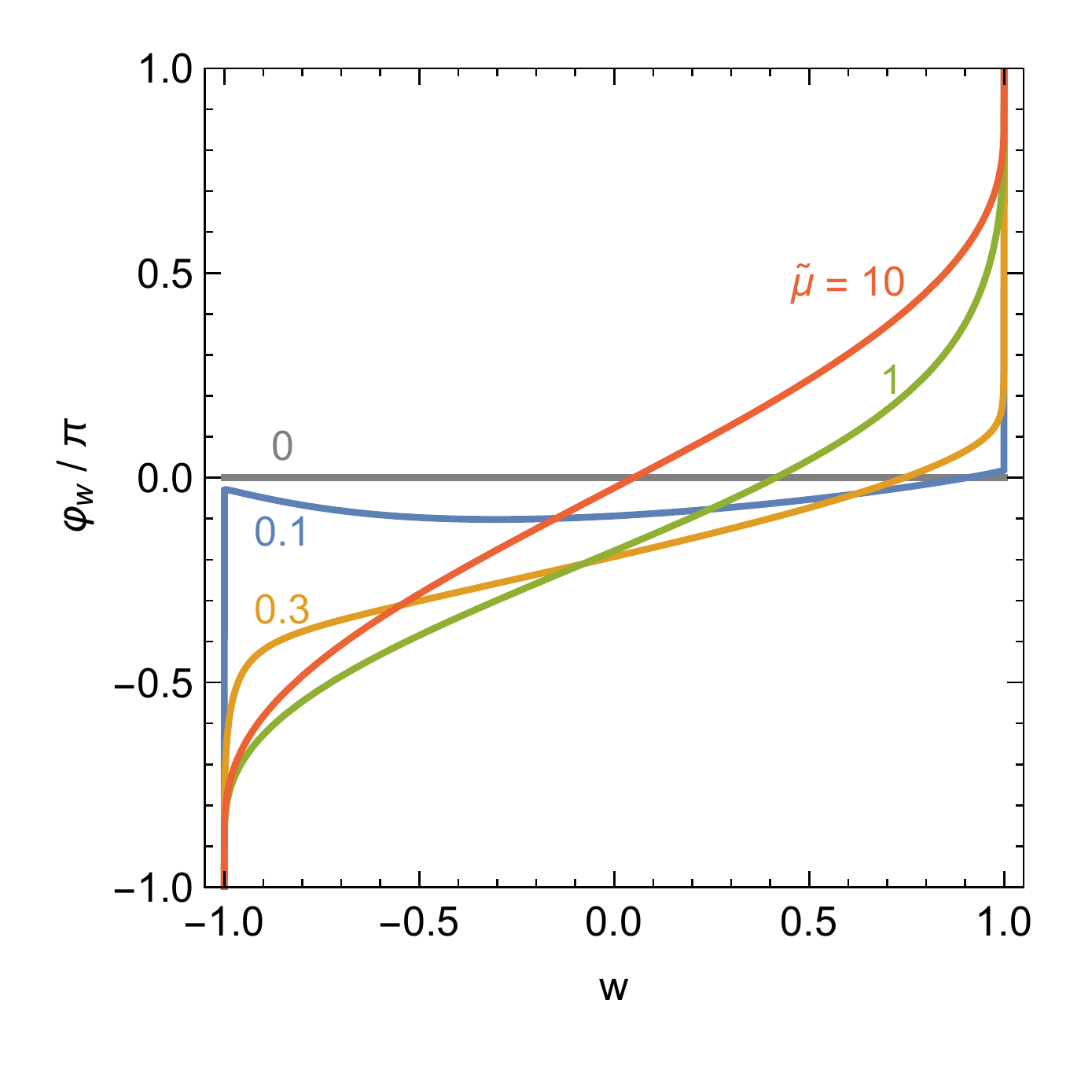}\hfil}
  \caption{Eigenfunction parameters for non-collective modes for an isotropic system
  for the indicated values of $\tilde\mu$.
  {\em Left panel:\/} Constant term $b_w$ according to equation~\eqref{eq:B-normalisation}.
  {\em Right panel:\/} Mixing angle $\varphi_w$  according to equation~\eqref{eq:phaseshift-2}.}
  \label{fig:isotropic-parameters}
\end{figure}

For an isotropic system we show $\varphi_w$ and $b_w$ for several
values of $\tilde\mu$ in figure~\ref{fig:isotropic-parameters}. The
mixing angle varies in the range $-\pi<\varphi_w<+\pi$ on the interval
\hbox{$-1<w<+1$}.  In the isotropic case, the Hamiltonian in
equation~\eqref{eq:Hamiltonian} is symmetric and thus Hermitean, so
the eigenfunctions must be orthogonal. It is a simple exercise to show
that indeed $\dashint_{-1}^{+1} du\,Q_{w_1,u}Q_{w_2,u}=0$ for
$-1<w_1<w_2<+1$ if we use the expressions $\alpha_w$ and $\beta_w$ for
the isotropic case given in equation~\eqref{eq:alpha-isotropic}. In addition,
the non-collective eigenfunctions are also orthogonal to the two collective
ones. For a non-isotropic system, the eigenfunctions are linearly independent,
but not orthogonal, so an explicit cross-check is less obvious.

\subsection{Discrete angles}

\begin{figure}[ht]
  \hbox to\textwidth{\hfil
  \includegraphics[scale=0.6]{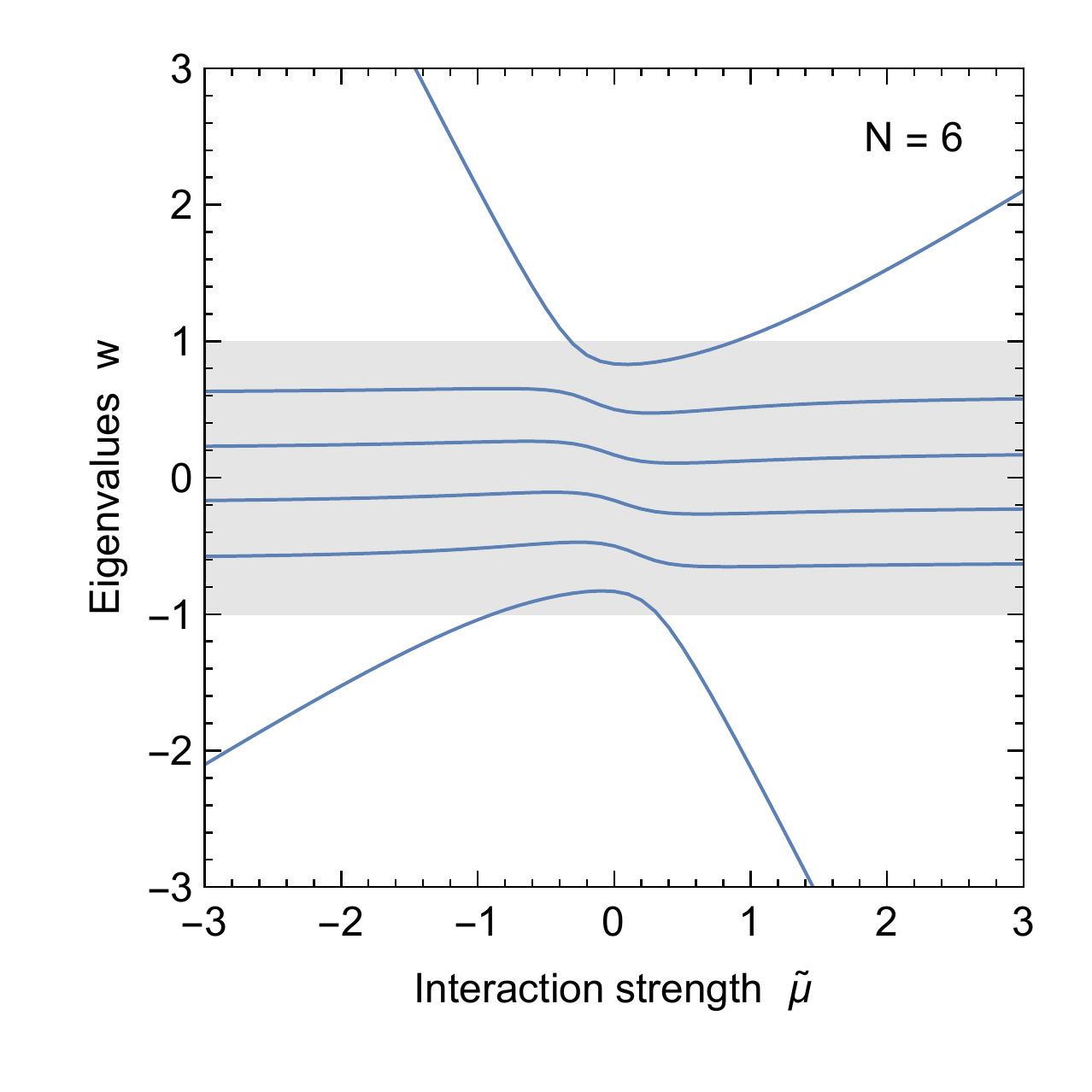}\hskip20pt
  \includegraphics[scale=0.6]{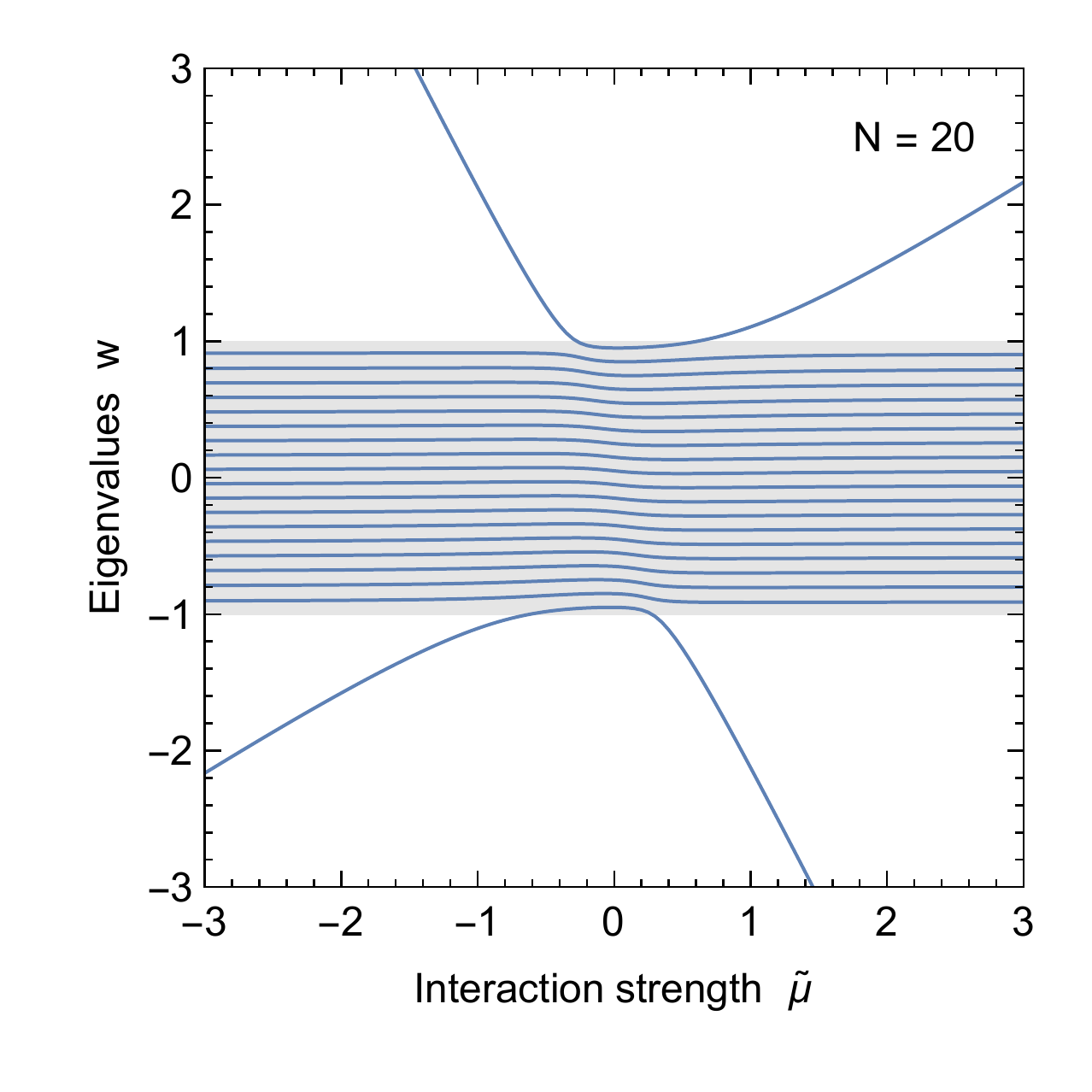}\hfil}
  \caption{Eigenvalues for the isotropic angle distribution in analogy to
  the right panel of
  figure~\ref{fig:isotropic-dispersion} for a discrete $u$ grid
  with $N=6$ and 20 grid points.}
  \label{fig:discrete-iso}
\end{figure}

It is instructive to solve the Schr\"odiger-like equation~\eqref{eq:EOM8b}
on a discrete grid of angles. For numerical solutions, this would have to be done
anyway and we already had to take conceptual recourse to discrete angles
to make sense of the continuum eigenfunctions and their normalisation.
Therefore, we may determine the eigenvalues and eigenvectors of the
``Hamilton operator'' $\cH$. In the discretised case it is an $N{\times}N$
matrix
\begin{equation}\label{eq:H-matrix}
  \sH_{ij}=\delta_{ij} u_i -\tilde\mu (1-u_iu_j)\,G_{j}\Delta u\,,
\end{equation}
without summation of repeated indices on the rhs, where
the angular spectrum is $G_i=G(u_i)$. The grid points are
\begin{equation}
  u_i=-1-\frac{\Delta u}{2}+i\,\Delta u
  \quad\hbox{for}\quad i=1,\ldots,N
\end{equation}
so each represents the center of a bin of width $\Delta u$. In the noninteracting
case the eigenvalues are $w_i=u_i$ and the normalised eigenvectors are
$Q_{i,j}=\delta_{ij}$, which is the $j$th component of the $i$th eigenvector.
In figure~\ref{fig:discrete-iso} we show the eigenvalues as a function
of $\tilde\mu$ for the isotropic system, in analogy to the right panel
of figure~\ref{fig:isotropic-dispersion} where the horizontal lines
were examples for the continuous distribution of eigenvalues, whereas
here we show solutions of the discrete system with $N=6$ and 20 grid
points. So here the eigenvalues $w_i$ are shifted relative
to the undisturbed ones $u_i$. Even for a small number of bins,
the two collective modes are very close to those found from the continuous
solution. The two collective modes develop from the two
modes at the edge of the light cone and peel off for increasing
interaction strength.

While the shift of the eigenvalues may seem small for large $N$, this is not
the case if we write them in the form
\begin{equation}
  w_i=u_i+\epsilon_i \Delta u\,.
\end{equation}
In units of the splitting $\Delta u$ between eigenvalues, the shift is not small
and approaches an asymptotic value for large $N$ for a fixed value of
$i/N$.

By the same logic that allowed us to determine the form of the eigenfunctions in the
continuum case, we find that the discrete eigenvector $k$ has components
\begin{equation}\label{eq:vec-1}
  Q_{k,i}=\left[\frac{A_k}{w_k-u_i}+B_k\right]\Delta u
  =\left[\frac{A_k}{(k-i+\epsilon_k)\Delta u}+B_k\right]\Delta u\,.
\end{equation}
They are normalised as $\sum_{i=1}^{N}Q_{k,i}^2=1$. This sum is convergent
because $\epsilon_k\not=0$, so the eigenvalues usually do not fall on
exact grid points. They do fall on grid points in the noninteracting case
and can do so also for special values of $\tilde\mu$, in which case
the eigenvector is simply $Q_{k,i}=\delta_{ki}$.
Starting from equation~\eqref{eq:vec-1} we can make the transition
to the continuous case with $\Delta u\to 0$, where the main trick is to
split the term proportional to $A_k$ in the form
\begin{equation}\label{eq:A-split}
  \frac{1}{k-i+\epsilon_k}=\frac{k-i}{(k-i)^2-\epsilon_k^2}-\frac{\epsilon_k}{(k-i)^2-\epsilon_k^2}\,.
\end{equation}
In the continuous limit, the first term in a summation turns to the principal-part
integration of $1/(w-u)$, whereas the second term becomes $\delta(w-u)$.
An integral over $u$ turns to a summation over $i$, so in the second term we need to
evaluate
\begin{equation}\label{eq:summation}
  \sum_{i=0}^{N}\frac{-\epsilon_k}{(k-i)^2-\epsilon_k^2}= \sum_{j=-k}^{N-k}\frac{-\epsilon_k}{j^2-\epsilon_k^2}
  \to  \sum_{j=-\infty}^{+\infty}\frac{-\epsilon_k}{j^2-\epsilon_k^2}
  =\pi\,{\rm cotan}(\pi\epsilon_k)\,,
\end{equation}
where we have extended the sum to $\pm\infty$ in the continuum limit $\Delta u\to 0$.

So more precisely, the second term in equation~\eqref{eq:A-split} turns to $\pi\,{\rm cotan}(\pi\epsilon_w)\,\delta(w-u)$.
Multiplying both terms with $\sin(\pi\epsilon_w)/\pi$ reveals  $\sin(\pi\epsilon_w)/[\pi(w-u)]+\cos(\pi\epsilon_w)\delta(w-u)$ for
the relative contributions
as in equation~\eqref{eq:phaseshift}. We conclude that
$\varphi_w=\pi\epsilon_w$, so the mixing angle between the two singular functions
is equivalent, in the discrete case, to the offset
$\epsilon_i$ between the grid points $u_i$ and the eigenvalues $w_i$ or equivalently
to the shift of the eigenvalues relative to the ones of the noninteracting system.

\section{Crossings of the angle distribution}

\label{sec:crossings}

\subsection{Single crossing}

\label{sec:single-crossing}

\begin{figure}[ht]
  \hbox to\textwidth{\hfil
  \includegraphics[scale=0.5]{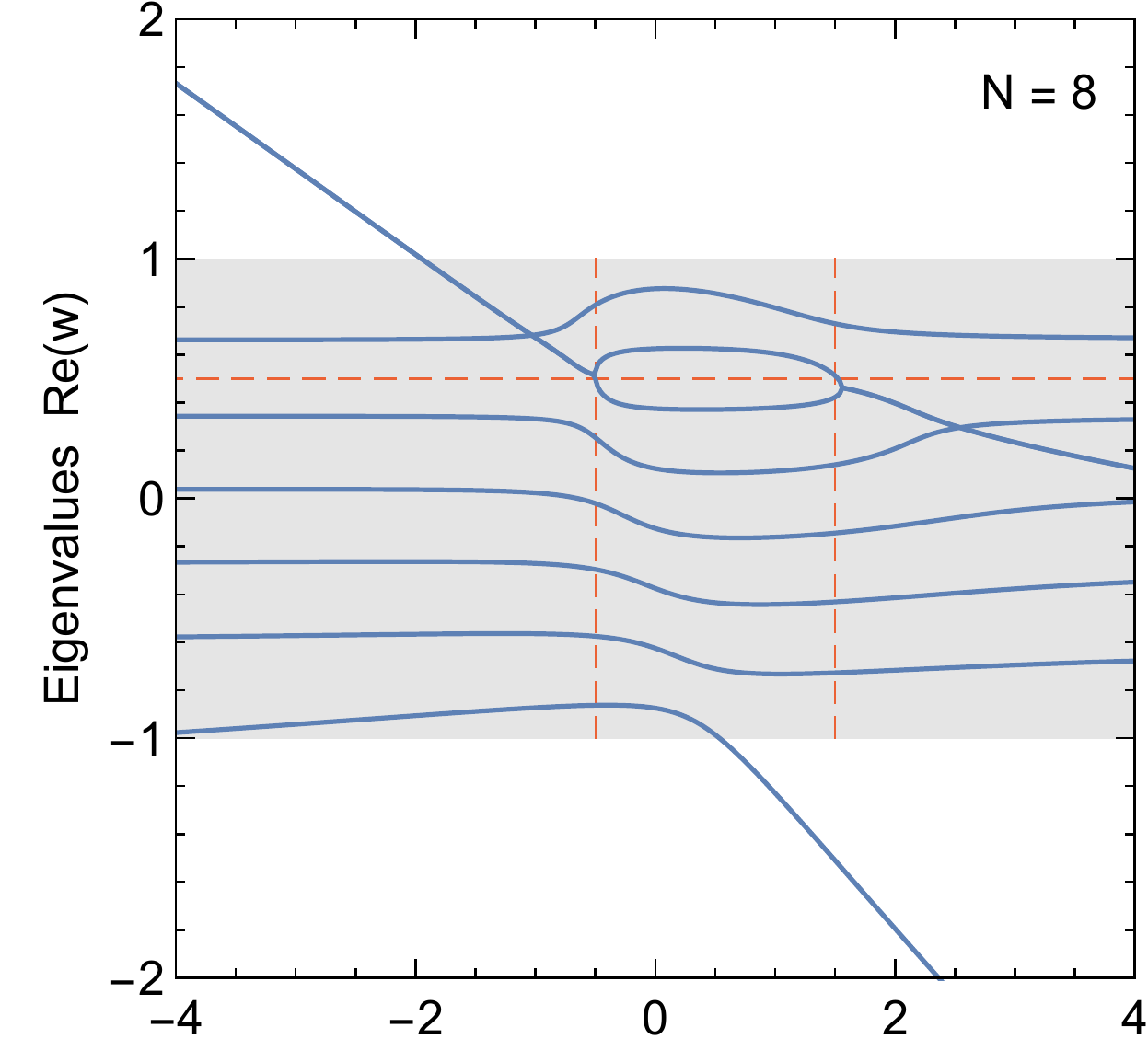}\hskip20pt
  \includegraphics[scale=0.5]{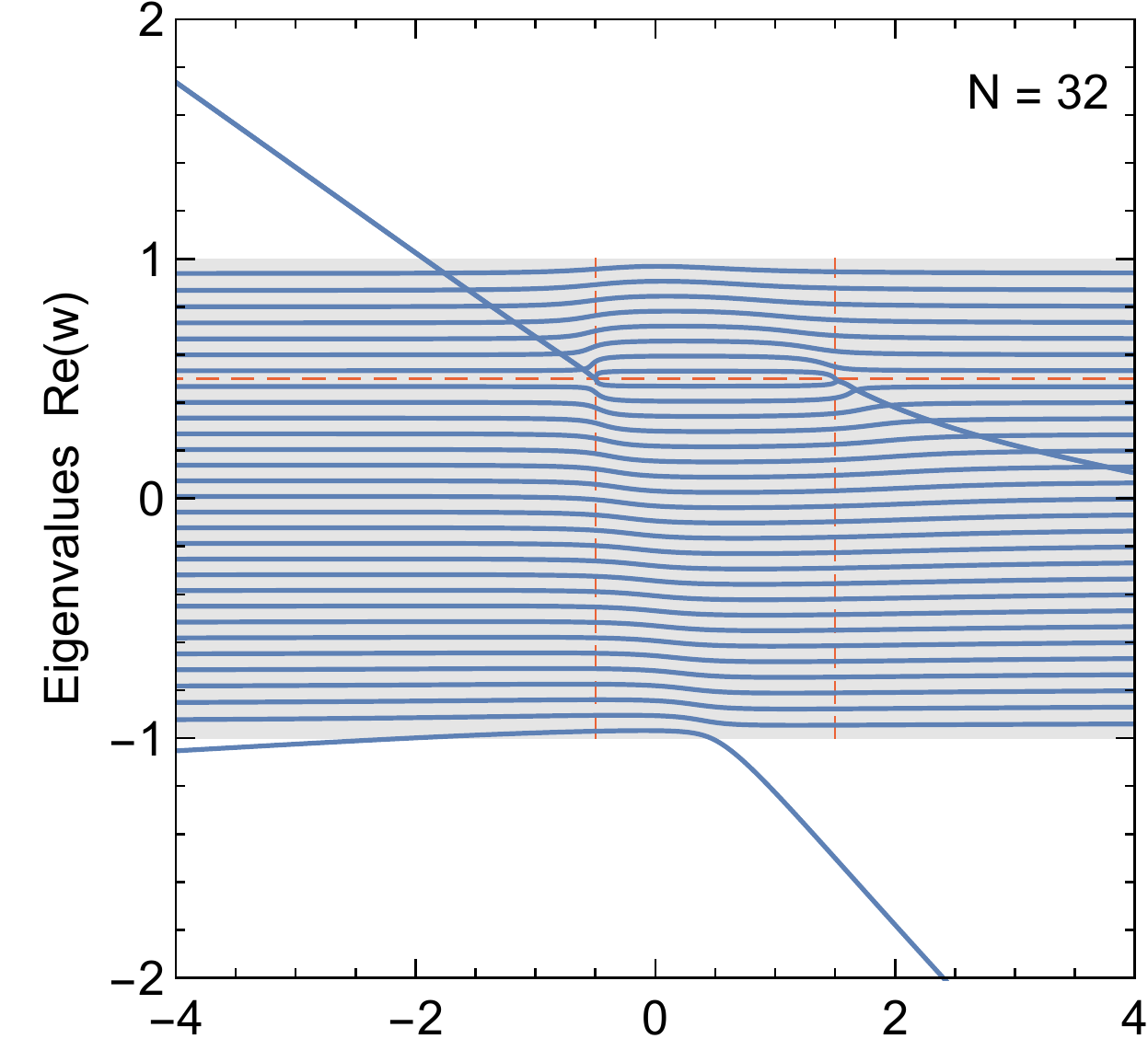}\hfil}
  \vskip2pt
  \hbox to\textwidth{\hfil
  \includegraphics[scale=0.5]{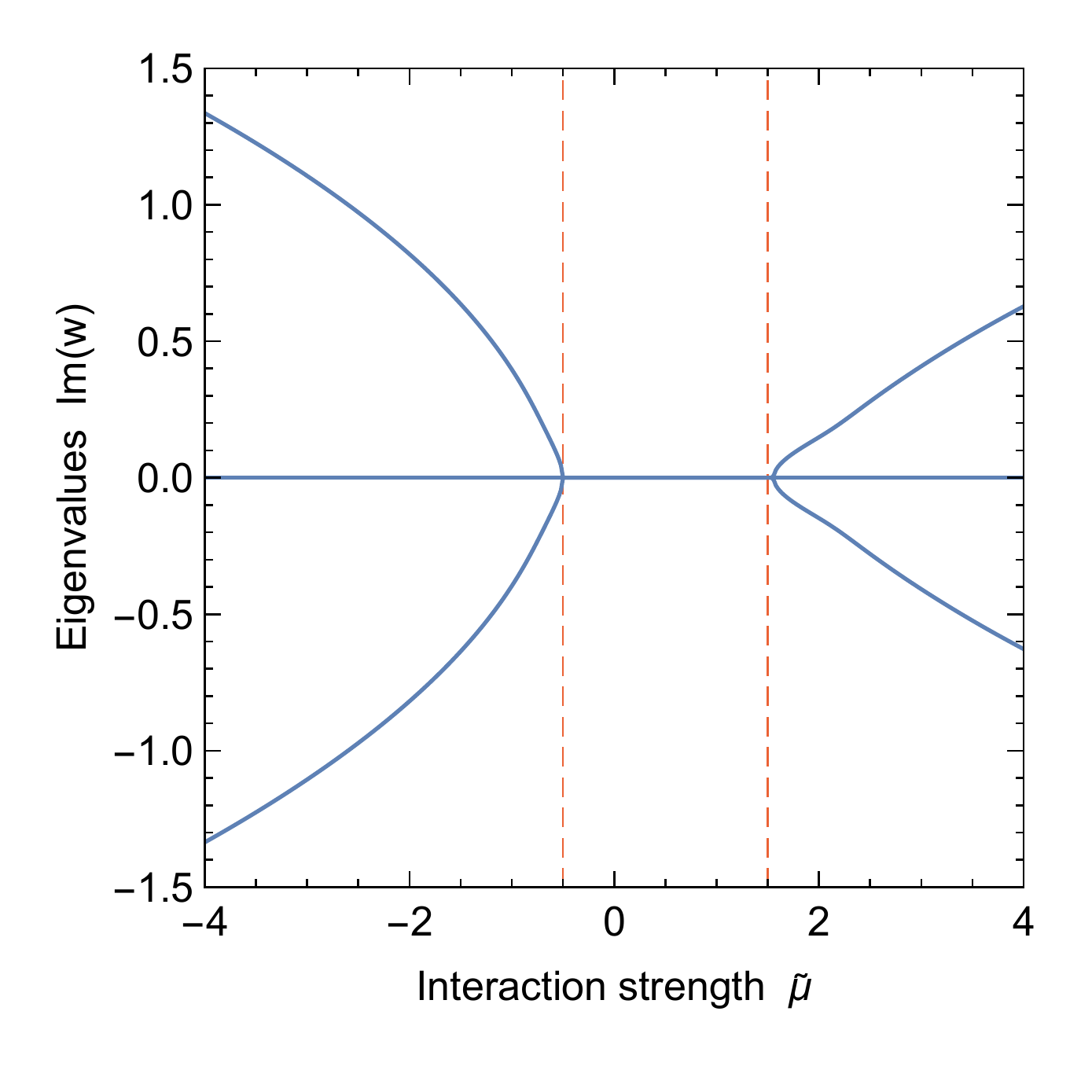}\hskip20pt
  \includegraphics[scale=0.5]{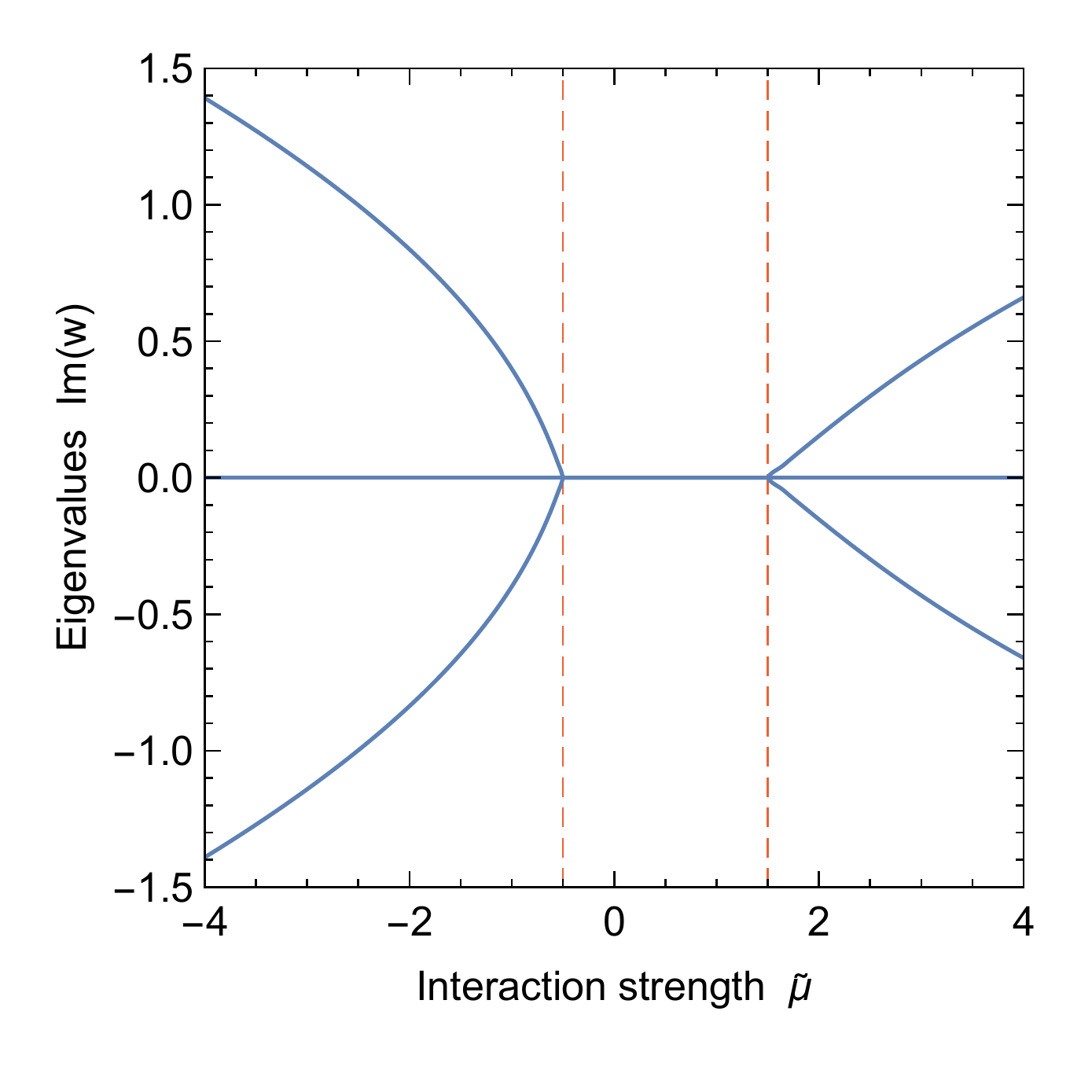}\hfil}
  \caption{Eigenvalues for a simple crossed spectrum as described in the text.
  {\em Top panels:\/} Real part. {\em Bottom panels:\/} Imaginary part. 
  {\em Horizontal dashed red line:\/} position of the crossing.
  {\em Vertical dashed lines:\/} predicted branching points for $N\to\infty$.}
  \label{fig:simple-cross}
\end{figure}

For a general non-isotropic function $G_u$, the solutions are
qualitatively similar to what we have shown so far if $G_u$ does not
change sign on the interval $-1<u<+1$. However, in the presence of a
crossing, tachyonic solutions can appear, which here mean complex
eigenvalues of the Hamiltonian $\cH$.  We first consider a simple
example of a schematic singly-crossed spectrum of the form
$G_u=\half-u$ that varies from $\frac{3}{2}$ to $-\half$ on the
interval $-1\leq u\leq+1$. We show the discrete eigenvalues in
figure~\ref{fig:simple-cross} for $N=8$ and 32 for both real and
imaginary part. We find both a real and a complex collective
solution. As observed earlier, the collective solutions depend only
mildly on the number of grid points.  Fast flavor oscillations do not
produce spurious instabilities that plague the numerical analysis of
slow flavor oscillations.

To identify the points where the complex solutions branch off within
the light cone, we note that for $\tilde\mu=0$ the only solutions are
non-collective ones. The branch point appears for increasing
$\tilde\mu$ and requires the merging of two real eigenvalues into a
pair of complex conjugate ones. In the continuous case, every real
$-1<w<+1$ is an eigenvalue with a unique eigenfunction determined by
the value of $\alpha_w$ that follows from the determinant
equation~\eqref{eq-alpha}.  The only exception is the location $w=\uc$
where the function $G(u)$ has a crossing and thus $G_w=G_{\uc}=0$ so
that $\alpha_{\uc}$ is not determined. Therefore, the branch point
must occur for $w=\uc$. In our example, this is for $w=\uc=1/2$, shown
as a horizontal dashed line in the upper panels of
figure~\ref{fig:simple-cross}.

At the branch point we need two eigenfunctions for the two degenerate
eigenvalues which merge at $w=\uc$ for some value of $\tilde\mu$. This
is only possible if the rhs of equation~\eqref{eq-alpha} vanishes,
in which case this equation is fulfilled because both sides are
separately zero. So the interaction strength for the branch points
follows from
\begin{equation}\label{eq:mu-crossing-1}
   F_{\uc}
  +\frac{1+\tilde\mu\cG_0(\tilde\mu\cG_0+\uc)+\tilde\mu\cG_1}{\tilde{\mu}\,
    (1-\uc^2-\tilde\mu \uc\cG_0+\tilde\mu\cG_1)}=0\,.
\end{equation}
This is a quadratic equation for $\tilde\mu$ with the solutions
\begin{equation}\label{eq:mu-crossing-2}
  \tilde\mu_{\rm c}=-\frac{2}{F_{\uc}(1-\uc^2)+\cG_0\uc+\cG_1
  \pm\sqrt{\bigl[F_{\uc}(1-\uc^2)+\cG_0\uc+\cG_1\bigr]^2-4\bigl[\cG_0^2+F_{\uc}(\cG_1-\cG_0\uc)\bigr]}}.
\end{equation}
In our example we find $\tilde\mu_{\rm c}=-1/2$ and $3/2$, shown
as vertical dashed lines.

This condition becomes more transparent if we write the angle spectrum, assumed to
have a crossing at $u=\uc$, in the form
\begin{equation}\label{eq:G-with-crossing}
  G(u)=(u-\uc)\,P(u)\,,
\end{equation}
where $P(u)$ is some other function that could have crossings at other
values of $u$. In analogy to equation~\eqref{eq:G-moments} we introduce
the moments of the $P(u)$ distribution
\begin{equation}\label{eq:P-moments}
  \cP_n=\int_{-1}^{+1}du\,P_u u^n\,.
\end{equation}
This definition implies $F_{\uc}=-\cP_0$ and
$\cG_n=\cP_{n+1}-\uc\cP_n$.  The interaction strength at the branching
points is then given entirely by the $P$-moments as
\begin{eqnarray}\label{eq:mu-crossing-3}
  \tilde\mu_{\rm c}&=&\frac{2}{\cP_0-\cP_2
  \pm\sqrt{\bigl(\cP_0+\cP_2+2\cP_1\bigr)\bigl(\cP_0+\cP_2-2\cP_1\bigr)}}
  \nonumber\\[1ex]
  &=&\frac{2}{\cP_0-\cP_2
  \pm\sqrt{\bigl(\cP_0+\cP_2\bigr)^2 -4\cP_1^2}}
  \,.
\end{eqnarray}
It is remarkable that $\tilde\mu_{\rm c}$
does not depend on the eigenvalue $w_{\rm c}$, i.e.,  $\tilde\mu_{\rm c}$
depends only on the first moments of the function $P(u)$. An alternative
derivation will be presented in appendix~\ref{app:CritPoints}.

This result becomes yet clearer if we introduce averages
$\llangle f(u)\rrangle=\int_{-1}^{+1}du\,P(u)\,f(u)$
that imply
\begin{equation}\label{eq:mu-crossing-4}
  \tilde\mu_{\rm c}=\frac{2}{\llangle1- u^2\rrangle
  \pm\sqrt{\llangle(1+u)^2\rrangle\llangle(1-u)^2\rrangle}}\,.
\end{equation}
If the function $G(u)$ has exactly one crossing, then the function $P(u)$ has
no crossing at all, implying that $\llangle(1+u)^2\rrangle$
and $\llangle(1-u)^2\rrangle$ has the same sign because
the integrand $(1\pm u)^2>0$. So the discriminant is positive
and $\tilde\mu_{\rm c}$ is real. We conclude that a single crossing indeed
leads to branching points and thus to eigenmodes with complex $\Omega$.

The crossing at $u=\uc$ could be of the form $G(u)=(u-\uc)^n Q(u)$ with
positive integer $n$. If $n$ is odd, the previous argument does not change
except that $P(u)$ would be non-negative instead of positive (or non-positive
instead of negative) and thus would have no crossing. For even $n$
there would be no crossing of $G(u)$ because it would not change sign
at $\uc$. In other words, it is enough that $G(u)$ changes sign at
$\uc$, the change need not be a simple crossing of the form $u-\uc$.

\subsection{Multiple crossings}

In principle, the function $G_u$ could change sign in several
places. For $n$ crossings it can then be written in terms of some
other function $P(u)$, assumed to have no crossings, as
\begin{equation}
G(u)=P(u)\,\prod_{j=1}^n(u-u_j)\,,
\end{equation}
assuming also $u_1<u_2<\ldots<u_n$.

A simple example with two crossings is $G(u)=(u-u_1)(u-u_2)$ with
\hbox{$-1<u_1<u_2<+1$}. Possible critical points are at $w_{\rm
  c}=u_{1,2}$. Considering $w_{\rm c}=u_1$, the discriminant
in equation~\eqref{eq:mu-crossing-4} is $\frac{16}{9}(4u_2^2-1)$
and thus positive for $|u_2|>\frac{1}{2}$. Everything is symmetric
under $1\leftrightarrow 2$, so we have two instabilities if
$|u_{1,2}|>\frac{1}{2}$, one instability if $|u_j|>\frac{1}{2}$
for $j=1$ or 2, but not both, and no instability at all if
$|u_{1,2}|<\frac{1}{2}$. This simple example proves that a
multi-crossed spectrum need not have any tachyonic solutions.

For a similar example with three crossings of the form
$G(u)=(u-u_1)(u-u_2)(u-u_3)$ we find cases with 1, 2 or
3 instabilities, depending on the values of $u_{1,2,3}$,
but there is always at least one instability.

We suspect that an odd number of crossings would guarantee at least
one unstable solution so that it would be enough for $G_u$ to have
opposite signs at $u=\pm1$. This would then be a more general
definition of what we mean with a crossing of the spectrum. However,
we have not pursued the question of how to prove this conjecture.

\section{Collective motion vs.\ dissipation}

\label{sec:decoherence}

\subsection{General expressions}

For the question of flavor conversion in an astrophysical environment with
dense neutrinos, the existence of unstable solutions is the main question
that can be addressed with a linearised stability analysis. If the initial
conditions provide a seed for an exponentially growing solution, it will eventually
dominate and push the system into the nonlinear regime.
However, here we are interested in the role of non-collective modes for
conditions without an exponentially growing solution and ask what
will happen to a nonvanishing initial condition of flavor coherence. This
might be a wave packet, but we here consider an even more elementary
case of a plane wave with wave vector $\bK$.

So effectively we consider solutions of the Hamiltonian
equation~\eqref{eq:hamiltonian-1} with a chosen interaction strength~$\tilde\mu$.
One way of defining the overall flavor coherence of the ensemble is by simply averaging over all modes
\begin{equation}
S_{\rm tot}(t)=\half\int_{-1}^{+1}du\,S_u(t)\,.
\end{equation}
The initial condition $S_u(0)$ can be expanded in the normalised eigenfunctions
$Q_{w,u}$, each contributing an amplitude $T_w$ such that
\begin{equation}\label{eq:expansion-1}
  S_u(0)=\int_{-1}^{+1}dw\,T_w\,Q_{w,u}+\sum_{j=1,2}T_{j}\,Q_{w_j,u}\,,
\end{equation}
where $j=1,2$ refers to the two collective eigenmodes with $|w_j|>1$, while
the non-collective modes are indexed with their eigenfrequency $-1\leq w\leq+1$.
Therefore, the overall flavor coherence evolves as
\begin{equation}
S_{\rm tot}(t)=\int_{-1}^{+1}dw\,T_w\,\overline Q_{w}\,e^{-i w t}
+\sum_{j=1,2}T_{j}\,\overline Q_{w_j}\,\,e^{-i w_j t}\,,
\end{equation}
where
\begin{equation}
\overline Q_{w}=\half\int_{-1}^{+1}du\,Q_{w,u}
\end{equation}
for both collective and non-collective modes.

In the noninteracting case, there are no collective modes and the
eigenfunctions of the non-collective ones are $Q_{w,u}=\delta(w-u)$,
where we use $A=1$ for the global normalisation factor.
Assuming $S_u(0)=1$ implies $T_w=1$ and $\overline{Q}_w=G_{w}$ so that
\begin{equation}\label{eq:decoherence-nonint}
  S_{\rm tot}(t)=\half\int_{-1}^{+1}dw\,e^{-i wt}=\frac{\sin(t)}{t}\,.
\end{equation}
As expected, the flavor coherence dissipates as $t^{-1}$
by kinematical decoherence
(de-phasing) of the continuum of energy eigenstates $w$.

\subsection{Isotropic system}

In the interacting case, the Hamiltonian of equation~\eqref{eq:Hamiltonian}
is usually not Hermitean and the eigenfunctions not orthogonal, so determining
the amplitudes $T_w$ requires the dual basis. The one exception is the
isotropic system ($G_u=1$), where the Hamiltonian is symmetric
and thus the eigenfunctions are orthogonal as discussed in
section~\ref{sec:normalisation}, so for simplicity we use this case
to illustrate the transition from decohering to collective behavior. In addition
we assume $S_u(0)=1$, implying for the non-collective modes
\begin{equation}\label{eq:Tw}
  T_w=\int_{-1}^{+1}du\,Q_{w,u}
\qquad\hbox{and}\qquad
  \overline{Q}_w=\half\int_{-1}^{+1}du\,Q_{w,u}=\half T_w
\end{equation}
and similar for the collective ones, so
\begin{equation}\label{eq:Stot}
S_{\rm tot}(t)=\half\int_{-1}^{+1}dw\,T_w^2\,e^{-i w t}
+\half\sum_{j=1,2}T_{j}^2\,e^{-i w_j t}\,.
\end{equation}
For the non-collective modes, we use the eigenfunctions
of the first line of equation~\eqref{eq:phaseshift} with the
coefficients $\alpha_w$ and $\beta_w$ for the isotropic case
given in equation~\eqref{eq:alpha-isotropic}, providing
\begin{equation}\label{eq:Tw-isotropic}
  T_w^{-2}=\pi^2\tilde\mu^2\left(2\tilde\mu w+w^2-1\right)^2
   +\left[1+2\tilde\mu(2\tilde\mu+w)+\tilde\mu\left(2\tilde\mu w+w^2-1\right)\log\left(\frac{1-w}{1+w}\right)\right]^2.
\end{equation}
For vanishing $\tilde\mu$ this is $T_w^2=1$ and we recover
equation~\eqref{eq:decoherence-nonint}. For large $\tilde\mu$, one finds
\begin{equation}\label{eq:Tw-large-mu}
  T_w^2=\frac{1}{4\tilde\mu^4}\,\frac{1}{\pi^2 w^2+\left[2+w\log\left(\frac{1-w}{1+w}\right)\right]^2}\,,
\end{equation}
so in this limit the contribution of the non-collective modes decreases with $\tilde\mu^{-4}$
and the evolution becomes entirely collective.

\begin{figure}[ht]
   \centering
  \includegraphics[scale=0.6]{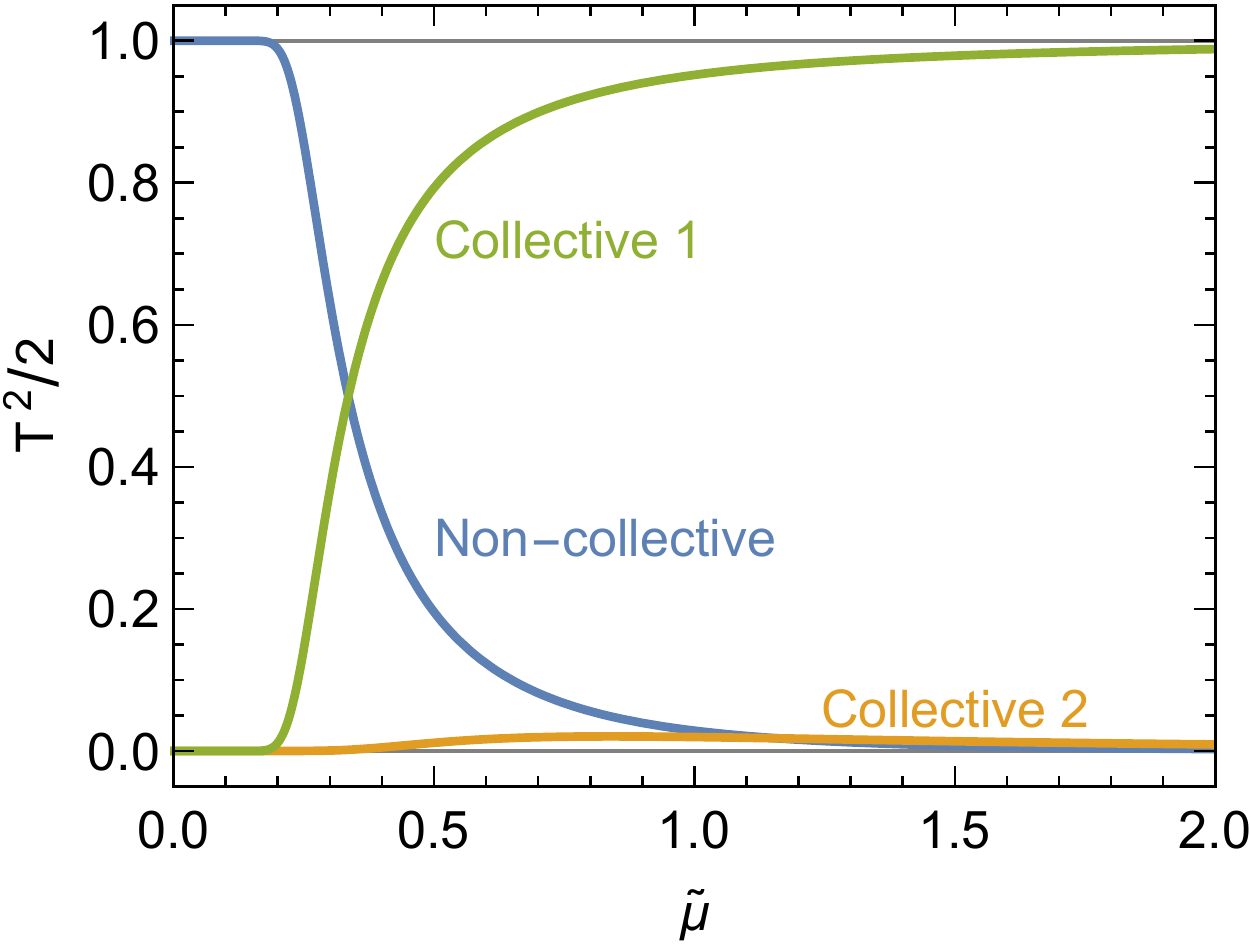}
  \caption{Contributions to $S_{\rm tot}(t)$ in equation~\eqref{eq:Stot}
  at $t=0$ from non-collective modes and the two collective ones. The three contributions
  add up to unity.}
  \label{fig:decoherence}
\end{figure}

The contributions $T_{1,2}^2$ of the collective modes are straightforward to calculate
from the collective eigenfunctions and eigenvalues, but the expressions become cumbersome.
In figure~\ref{fig:decoherence} we show  $\half T_{1,2}^2$ as a function of $\tilde\mu$
as well as the total contribution from the non-collective modes. The three contributions add up
to unity. In our example of an isotropic medium, the collective mode No.~1 has the eigenfunction
$Q_{1,u}=1/\sqrt{2}$ in the large-$\tilde{\mu}$ limit, identical to the initial condition
of $S_u$, leaving no projection on the other modes, so that for  large $\tilde{\mu}$
only one eigenfrequency contributes.
(The other collective mode has the large-$\tilde{\mu}$ eigenfunction $Q_{2,u}=\sqrt{3/2}\,u$
and thus has no overlap with a constant function on the interval $-1<u<+1$.)

We conclude that for $\tilde\mu\lesssim0.3$, the initial flavor coherence
dissipates by decoherence, whereas for somewhat larger interaction strength
one of the collective modes takes over. For an initial wavepacket, the components
with $K$ exceeding a few times $\mu$ completely dissipate, whereas the smaller-$K$ components propagate
as collective flavor excitations. In an intermediate range, the initial flavor
coherence partially dissipates.

We mention in passing the
similarity to the question of synchronized oscillations
vs.\ decoherence at intermediate densities for ordinary collective
oscillatoins
(driven by the mass term) that was addressed in
reference~\cite{Raffelt:2010za}. In that case, the dephasing effect
came from different energies in $\Delta m^2/2E$, whereas the angle
degree
of freedom was intergrated out.

\section{Conclusions}

\label{sec:conclusions}

We have re-examined the dispersion relation for the mean field of
flavor coherence in a dense neutrino gas, considering in particular
``fast modes,'' where neutrino masses do not enter and the
question of which modes are supported depends on the
neutrino angle distribution, or rather, on the angle distribution of
the electron lepton number (ELN) carried by neutrinos.  The purpose of
our discussion was primarily to illuminate the connection between
non-collective modes that exist even in the absence of
neutrino-neutrino interactions and collective modes that appear as a
result of neutrino refraction in a background of dense neutrinos.

In the absence of collective effects, the angle-dependent neutrino
flux simply carries along any putative flavor coherence that may have
been imprinted initially. Unless we consider very special angle
distributions, such as two-beam toy models that have sometimes been
used in the literature, the angular range of neutrino directions will
quickly decohere any initially prepared flavor coherence.

The presence of neutrino-neutrino interactions causes the appearance
of collective modes that may or may not show tachyonic instabilities,
whereas the non-collective modes get modified. We have derived
explicit expressions for the non-collective eigenfunctions and we have
also shown that, as the interaction strength is increased, the
amplitude of flavor coherence carried by these modes decreases. For a
large interaction strength, the emergent collective modes dominate the
evolution.

For fast flavor modes, the neutrino-neutrino interaction energy that
we have
defined as $\mu=\sqrt2\GF (n_{\nu_e}+n_{\bar\nu_e})$ is the only energy scale of the
problem, so all wave vectors and frequencies are naturally measured in
units of $\mu$. So for flavor modes with $K\gg\mu$ collective effects
are practically irrelevant, whereas for $K\lesssim\mu$ they are
important. Of course, for unstable modes, collective effects are
always important in the sense that any unstable mode eventually
dominates if it has enough time to grow.

We have also clarified that the appearance of
tachyonic collective modes can be understood as the coalescence of two
non-collective modes, i.e., two real eigenfrequencies merge at a
critical point to become two complex-conjugate solutions.
This happens at crossings of the neutrino
angle distribution and allows us to calculate the required interaction
strength. As a corollary we show that a single crossing of the
ELN angle distribution guarantees a tachyonic instability, whereas the
case of several crossings is less clear at present.

In this way the role of the non-collective and collective modes of the
linearised equations of motion of neutrino flavor coherence has been
further illuminated. Of course, these dispersion relations only allow
us to judge which kind of solutions are supported by a given dense
neutrino gas, but do not show which ones will actually be excited and
by what. Flavor coherence of the neutrino mean field probably can
be sourced ultimately only by the mass term, perhaps in conjunction
with density fluctuations of the medium. In this sense, understanding
the dispersion relation is probably better suited to understand if
tachyonic solutions are or are not supported by the neutrino gas, but
if they are actually excited, and with which strength, and how they
evolve in the nonlinear regime are questions that go beyond the topics
that can be addressed by the dispersion relation alone.


\section*{Acknowledgments}

We acknowledge partial support by the Deutsche Forschungsgemeinschaft through Grants
No.\ SFB 1258 (Collaborative Research Center {\em Neutrinos, Dark Matter,
Messengers}) and EXC 2094 (Excellence Cluster {\em Origins}), as well as the
European Union through Grant
No.\ H2020-MSCA-ITN-2015/674896 (Innovative Training Network
{\em Elusives}).

\appendix

\section{Critical points: Alternative derivation}
\label{app:CritPoints}

We can derive the critical points $(w_{\rm c},\tilde\mu_{\rm c})$
derived in section~\ref{sec:single-crossing}
where a tachynoic solution appears without having to worry about
the non-collective eigenfunctions. To this end we recall that
any collective mode fulfills equation~\eqref{eq:determinant-1}
which reads, with $\tilde\mu=\mu/K$,
\begin{equation}\label{eq:determinant-1a}
  \left\|\,\begin{pmatrix}
             1 &     \\
                &  -1
           \end{pmatrix}+\tilde\mu
\begin{pmatrix}
             \langle 1\rangle & \langle u\rangle   \\
             \langle u\rangle  & \langle u^2\rangle
           \end{pmatrix}
\right\|=0\,.
\end{equation}
Considering a complex eigenvalue, we now interpret $w$ as its real
part and $\kappa$ its imaginary part so that the definition
of equation~\eqref{eq:G-integrals} now reads explicitly
\begin{equation}\label{eq:G-integrals-a}
  \langle u^n\rangle=\int_{-1}^{+1}du\,G_u\,\frac{u^n}{w+i\kappa-u}\,.
\end{equation}
At the critical point, the imaginary part is infinitesimally small, so
we can use the Sokhotski-Plemelj Theorem in the form
\begin{equation}
  \int_{-1}^{+1}\,du\,\frac{f(u)}{w+ i\kappa-u} ~~\mathop{\longrightarrow}_{\kappa\to0}~~
  \dashint_{-1}^{+1}\,du\,\frac{f(u)}{\wc-u}- i\pi f(\wc)\,,
\end{equation}
where the integral on the rhs is the Cauchy principal
value. Therefore, in equation~\eqref{eq:determinant-1a} we should insert
\begin{equation}\label{eq:G-integrals-b}
  \langle u^n\rangle \to p_n-i\pi\,G_{\wc} \wc^n
\qquad\hbox{with}\qquad
  p_n=\dashint_{-1}^{+1}du\,G_u\,\frac{u^n}{\wc-u}\,.
\end{equation}
Equation~\eqref{eq:determinant-1a} then must be fulfilled
simultaneously for the real and imaginary part of the determinant,
the real part implying
\begin{equation}\label{eq:mu-crossing-3a}
  \tilde\mu_{\rm c}
  =\frac{2}{p_0-p_2\pm\sqrt{(p_0+p_2)^2 -4p_1^2}}\,.
\end{equation}
The imaginary part vanishes if
\begin{equation}\label{eq:mu-crossing-3b}
  G_{\wc}\left[\wc^2-1+\mc\left(p_2-2p_1\wc+p_0\wc^2\right)\right]=0\,.
\end{equation}
One solution is $G_{\wc}=0$ which happens exactly at the crossings of
$G_u$, i.e., when $\wc=\uc$. If we express again $G(u)=(u-\uc)P(u)$,
then the quantities $p_n$, for $\wc=\uc$, correspond to $\cP_n$
defined in equation~\eqref{eq:P-moments}, so our
new equation~\eqref{eq:mu-crossing-3a} is identical with
the previous result of  equation~\eqref{eq:mu-crossing-3}.

Another solution of equation~\eqref{eq:mu-crossing-3b} is when the expression in
square brackets vanishes, providing another expression for $\mc$ that must
agree with equation~\eqref{eq:mu-crossing-3a}, in principle allowing us to calculate
an expression for $\wc$. Simple examples do not provide consistent solutions
of these equations. The discussion in section~\ref{sec:single-crossing}
suggests that critical points only appear at crossings of $G(u)$ where
$G(\wc)=0$, but no proof is available that there could not exist unforeseen special
cases of other situations.

\providecommand{\noopsort}[1]{}\providecommand{\singleletter}[1]{#1}%

\providecommand{\href}[2]{#2}\begingroup\raggedright\endgroup

\end{document}